\documentclass{iopart}
\pdfoutput=1 

\usepackage[pdftex]{color,graphicx,hyperref}

\expandafter\let\csname equation*\endcsname\relax
\expandafter\let\csname endequation*\endcsname\relax 

\usepackage{amsmath,amsfonts,amssymb}
\usepackage{enumerate}
\usepackage{epsfig}
\usepackage{bbding}
\usepackage{textcomp}
\usepackage{wasysym}

\usepackage{subfigure}
\usepackage{amsthm}

\usepackage{algorithmic}
\usepackage{algorithm}

\usepackage[table,dvipsnames]{xcolor} 
\usepackage{colortbl} 

\newcommand{\iet}{\ensuremath{\tau}}
\newcommand{\tauzero}[1][]{\ensuremath{\tau_{R #1}}}
\newcommand{\tauone}[1][]{\ensuremath{\tau_{R1 #1}}}
\newcommand{\tauinf}[1][]{\ensuremath{\tau_{R \infty #1}}}
\newcommand{\eventinf}{\ensuremath{f_{e}} }
\newcommand{\avg}[1]{\ensuremath{\overline{ #1 }}}

\makeatletter
\newcommand{\ensavg}[1]{\ensuremath{\langle #1\ensavgaux}}
\def\ensavgaux{\@ifnextchar[\ensavgaux@i\ensavgaux@ii}
\def\ensavgaux@i[#1]{\ensuremath{| #1 \rangle}}%
\def\ensavgaux@ii{\ensuremath{\rangle}}%
\makeatother

\begin{document}

\title{\bf Multiscale Analysis of Spreading in a Large Communication Network}
\author{Mikko~Kivel\"a$^1$\footnote{Corresponding author: mikko.kivela@aalto.fi}, Raj~Kumar~Pan$^1$, Kimmo~Kaski$^1$, J\'anos~Kert\'esz$^{1,2}$, Jari~Saram\"aki$^1$ and M\'arton~Karsai$^1$}

\address{$^1$ BECS, School of Science and Technology, Aalto University, P.O. Box 12200, FI-00076}
\address{$^2$ Institute of Physics and BME-HAS Cond. Mat. Group, BME, Budapest, Budafoki \'ut 8., H-1111}

\date{\today}

\begin{abstract}
In temporal networks, both the topology of the underlying network and the timings of interaction events can be
crucial in determining how some dynamic process mediated by the network unfolds. 
We have explored the limiting case of the speed of spreading in the SI model, set up such that an event between an infectious and susceptible individual
always transmits the infection. The speed of this process sets an upper bound for the speed of any 
dynamic process that is mediated through the interaction events of the network. With the help of temporal networks derived from large-scale time-stamped data on mobile phone calls, we extend earlier results  that point out the slowing-down effects of burstiness and temporal inhomogeneities. In such networks, links are not permanently active, but dynamic processes are mediated by recurrent events taking place on the links at specific points in time.
We perform a multi-scale analysis and pinpoint the importance of the timings  of event sequences on individual links, their correlations with neighboring sequences, and the temporal pathways taken by the network-scale spreading process. This is achieved by studying empirically and analytically 
different characteristic relay times of links, relevant to the respective scales, and a set of temporal reference models that allow for removing selected time-domain correlations one by one.
\end{abstract}

\vspace{2pc}
\noindent{\it Keywords}: Network dynamics, Communication, supply and information networks, Socio-economic networks, Epidemic modelling
\maketitle

\section{Introduction}
Dynamics on complex networks often take place in a non-continuous manner where interactions are not permanent. Temporal networks~\cite{TempReview} constitute the adequate framework for such a situation, where a link between two nodes is present only for the period of the interaction. When aggregated over time, such systems can be represented with static weighted networks~\cite{Barrat2004}, where the link weights measure averaged link activity. Some of the properties of dynamic processes taking place on networks depend strongly on static network characteristics, while a detailed analysis of the temporal aspect could 
lead to further important insights.

Spreading is one of the most important processes in complex systems. It is relevant for a number of fields and applications 
ranging from epidemiology of biological viruses to the dynamics of social processes, such as opinion dynamics or information transmission~\cite{Boccaletti06}. While certain static characteristics of complex networks work to enhance spreading, such as the small world property, it has been shown that the temporal characteristics of links may slow it down~\cite{Vazquez07,Karsai11,Miritello10}. These results indicate that dynamic processes cannot necessarily take advantage of topologically shortest paths~\cite{Pan11}.
In addition, also static topological characteristics such as prominent community structure have been shown to give rise to  considerable decelerating effects on spreading speed~\cite{Simonsen04, Holme05, Karsai11}, while a fat-tailed degree distribution has been shown to be an accelerating property~\cite{Barthelemy04}. Furthermore, in weighted networks, the relationship between weights and topology provides an additional source of possible influence on the spreading dynamics. Especially for social networks it is known that links within communities are strong, while links
between them are weaker~\cite{Onnela07} -- such Granovetterian structure enhances the trapping effect by the 
communities, leading to further slowing down of spreading~\cite{Onnela07,Karsai11}. 

Spreading dynamics is typically studied using one of the standard epidemic models, where nodes are usually assigned to one of the three states -- (S)usceptible, (I)nfectious, and (R)ecovered~\cite{AndersonMay,Hethcote,Newman02,Kenah07}. The states that constitute 
the model are chosen depending on the problem at hand (e.g., SI, SIR, SIS). The models are then characterized by the rate of transmission between infectious and susceptible individuals upon contact, and the rate of recovery of infectious individuals. For static networks, transmission is usually assumed to take place with a probability uniform in time between infectious and susceptible network neighbors, \emph{i.e.}, it is 
assumed that the timings of contacts along links are determined by a Poisson process. 
However, in reality the timings of such contact sequences are heterogeneous and display several kinds of temporal correlations.  This is the case, e.g., in sexual contact networks~\cite{Morris97, RochaPNAS2010, Rocha2011} and human communication networks~\cite{Barabasi05, Vazquez07, Miritello10, Iribarren10, Karsai11}. For both cases, it has been shown that the spreading dynamics are considerably affected by the temporal features of the contact sequences.

In this article, we employ a deterministic version of the SI epidemic model, where susceptible nodes always become infected if they are in contact with infectious nodes through a contact event, \emph{i.e.}~a phone call. We restrict ourselves to a case where only one node is initially set infected at a random point in time.
Although simple, this model is useful because it gives an upper limit to the speed at which any dynamical process where nodes affect their neighbors through contacts can evolve. This includes all the other spreading processes.
The setting is the same as in our previous study~\cite{Karsai11}, where it was shown using temporal reference models that in addition to static structural features, heterogeneous contact sequences slow down spreading on empirical temporal networks of mobile telephone calls and emails.
We extend these results by organizing the reference model framework such that we can observe the hierarchical relationship between them. This allows us to
quantify the significance of both topological and temporal correlations to the dynamic spreading speed, and also to discover their relative importance.

We perform a multiscale analysis of the effects of temporal inhomogeneities on spreading dynamics, beginning with the role of individual links and moving then on to larger scales. We zoom in into the network and focus on the
effect of temporal inhomogeneities in the transmission speed of infection through single links in isolation, captured with the relay time of the link. 
Assuming random arrival of infection at one of the connected nodes enables to analytically calculate the effect of the inter-event time distribution \cite{Vazquez07}. It turns out that the expected transmission speed depends on the second moment of that distribution, or equivalently, on the 
burstiness of the sequence of events taking place on the link. However, such a simplified picture ignores correlations between such event sequences. These are taken into account in two steps. At the next scale, we consider relay times that take nearest-neighbor correlations into account with an approach similar to that  discussed by Miritello \emph{et al.}~\cite{Miritello10}. We build a model which shows that this type of correlations are likely to speed up the local spreading process.
Then, we make an attempt to take into account network-wide correlations of the event sequences and measure the actual speed of transmission of infection through the links during the global scale SI process. 
Our investigations show that most of the temporal inhomogeneities in the slowing down of spreading can be attributed to the single link properties, and although the static link weight dominates the time the infection waits before crossing the link, there is a high variation between waiting times of links with the same weight. The inclusion of the neighbors and the effect of the whole network both lower the relay time estimates.

We begin by introducing the data set and the reference models for the contact sequence that we are going to use in all of the following sections. In Section III, we study the spreading speed at the scale of the entire network. We explore the effects of different inhomogeneities, both dynamical and topological, and show that the bursts in the event sequences of 
links significantly slow down the spreading -- their effect is stronger than those of the weight-topology correlations or higher-order topological correlations such as communities. 
Finally, in Section IV we focus on the scale of links, defining the quantities 
that measure the spreading speed through individual links in isolation and in relationship to activation sequences by the other links.

\section{Data and methods}

\subsection{Data}
The data used in the following is based on the mobile phone call (MPC) records 
of seven million private customers of a single mobile network operator, covering
$\sim 20\%$ of the population of a European country. The data contain a sequence of
$\sim 444$ million time-stamped voice call events over a period of 120
days. In order to consider real social relations, we
only consider communication between pairs of subscribers who have at least one
reciprocated pair of calls between them. The temporal network, \emph{i.e.} the 
call event sequence used in the study is constructed as follows. First, we 
aggregate the events over the entire period to form a weighted network, 
where the nodes represent subscribers and a link between two nodes indicates
at least one call both ways. For static reference models, we also compute link weights
as the total number of calls made during the observation period.
 We then extract the
largest connected component (LCC) of the aggregated network; the LCC has
$N=4.6\times 10^6$ nodes and $L=9 \times 10^6$ links. The temporal network
employed in the study is then constructed as the time-stamped sequence of 
calls between all nodes in the LCC, and contains 
$306$ million call records. The static LCC has broad degree distribution 
and shows small-world properties. Its average degree
$\avg{k}=3.96$, and the average shortest path length between nodes
$\langle \ell \rangle=12.31$.  By geographic sampling of the network,
we observe that $\langle \ell \rangle$ varies logarithmically with the system
size. 
Since the observation period is finite, we apply periodic temporal boundary conditions
 by repeating the call sequence once its last event is reached.

\subsection{Reference models}
\label{sec:refmodels}
For static networks, a common way to assess 
the significance of chosen topological features is to compare their abundance or characteristics against some reference model where the network is randomized. This approach has also been applied in assessing the importance of such features for 
dynamical processes. The most widely applied reference model is the configuration model~\cite{Molloy95}, where the links of the original network are rewired pairwise randomly. 
This reference model preserves the original degree sequence but yields networks that are as random as the degree sequence allows. Then, one can assess the significance of topological characteristics of the empirical graph \emph{e.g.}~by measuring the extent to which the dynamics of some processes differ when they take place on the original networks or the reference ensemble. 

For temporal graphs, a similar approach can be applied: in this case, the original event sequences are randomized or randomly reshuffled to remove time-domain structure and correlations~\cite{Holme05,Karsai11}. Thus a reference model for an empirical event sequence is a maximally random ensemble of event sequences, for which some predefined set of properties are the same as for the empirical sequence. There are various  
kinds of temporal correlations, and thus no single, general-purpose reference model (a ``temporal configuration model'') can be designed. Rather, by applying appropriate reference models, one may switch off selected types of correlations in order to understand their contribution to the observed properties. 
The reference models used in this paper are listed in Table~\ref{table1} and a short explanation is given below. For full details, see \ref{sec:reference_models}.

\emph{The equal-weight link-sequence shuffled model.} Whole event sequences with time stamps are randomly exchanged between links that have the same weights, \emph{i.e.} numbers of events. Timing correlations between adjacent links are destroyed. While temporal characteristics of link event sequences are retained, any correlations between them and the topology are lost. All other temporal and structural correlations are retained.

\emph{Link-sequence shuffled model}. As above, but sequences are exchanged between links of any weight. Thus, weight-topology correlations are additionally destroyed.

\emph{Time-shuffled model}. The time stamps of the whole event sequence are randomly reshuffled. Thus all temporal correlations with the exception of network-level frequency envelope (for calls, the daily pattern~\cite{Jo11}) are destroyed, while all topological features are retained.

\emph{Uniformly random times}. Time stamps of all events are drawn uniformly from the observation period; all temporal correlations are destroyed.

\emph{Configuration model}. Topological correlations are destroyed with the configuration model that retains the degree sequence and keeps the network connected. Original event sequences of links are then randomly redistributed. Only single-link temporal characteristics remain. 

\emph{Random network}. A connected Erd\H{o}s-R\'enyi network is generated with the original number of nodes and links. Event sequences are distributed as above.

\begin{table}[h!]
\begin{center}
\begin{tabular}{|l||c|c|c||c|c|c||c|}
\hline
& \multicolumn{3}{c||}{Static} & \multicolumn{3}{c|}{Temporal} &\\ \cline{2-7}
\raisebox{1.5ex}[0PT]{Shuffling method} & DD           & C            & W            & D            & B            & E    &  \raisebox{1.5ex}[0PT]{Color}      \\ \hline\hline
Original                              & $\checkmark$ & $\checkmark$ & $\checkmark$ & $\checkmark$ & $\checkmark$ & $\checkmark$ &\cellcolor{Red}\\ \hline
Equal-weight link-sequence shuffled   & $\checkmark$ & $\checkmark$ & $\checkmark$ & $\checkmark$ & $\checkmark$ & \texttimes   &\cellcolor{OliveGreen}  \\ \hline
Time shuffled                         & $\checkmark$ & $\checkmark$ & $\checkmark$ & $\checkmark$ & \texttimes   & \texttimes   &\cellcolor{Blue} \\ \hline
Uniformly random times                & $\checkmark$ & $\checkmark$ & $\checkmark$ & \texttimes   & \texttimes   & \texttimes   &\cellcolor{RubineRed} \\ \hline
\hline
Link-sequence shuffled                & $\checkmark$ & $\checkmark$ & \texttimes   & $\checkmark$ & $\checkmark$ & \texttimes   &\cellcolor{SkyBlue} \\ \hline
Configuration model                   & $\checkmark$ & \texttimes   & \texttimes   & $\checkmark$ & $\checkmark$ & \texttimes   &\cellcolor{Goldenrod} \\ \hline
Random network                        & \texttimes   & \texttimes   & \texttimes   & $\checkmark$ & $\checkmark$ & \texttimes   &\cellcolor[gray]{0.5} \\ \hline
\end{tabular}
\end{center}
\caption{Properties that are retained in different random reference models. Descriptions of the abbreviations are given below. 
The properties which can be measured from an event sequence are in parenthesis.
DD - degree distribution (degree sequence), 
C - static configurations (adjacency matrix), 
W - weight-topology correlations (weight matrix), 
D - global time sequence (sorted list of all event times), 
B - event sequences in links (sorted list of even sequences on each link), 
E - everything (the whole event sequence).
The number of links and the link weight distribution of the empirical event sequence are retained in all reference models.
Further, we always require that reference networks are connected. The colors defined here for each reference model are used in the rest of the article.
}
\label{table1}
\end{table}

\section{Network-level spreading dynamics}
\label{sec:spreading}

\begin{figure}[th!] \centering
  \includegraphics[width=1.0\linewidth]{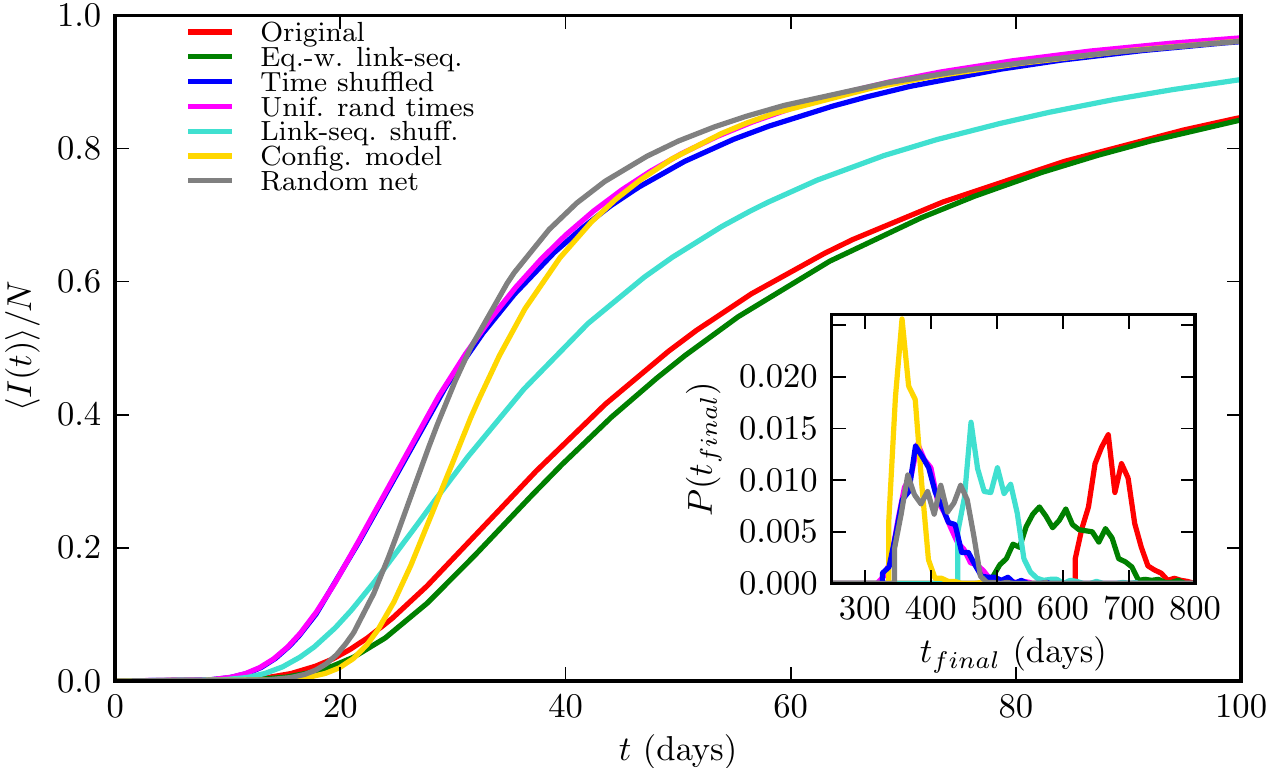}
\caption{The average fraction of infected nodes for SI spreading
dynamics on the MPC data and various reference models. (Inset) For each
model, the distribution of the final infection time $t_{final}$ is 
shown. All simulations were performed over $10^3$ initial conditions.}
\label{fig:SI_plots}
\end{figure}

We begin by simulating the SI spreading process on the empirical call
sequence data and with a set of six reference models. We set one random node's
state to infectious at a random point in time $t=t_0$, and let the information spread through the 
call events, measuring the time $t_{\mathrm{inf},i}$ from $t_0$ at which each of the 
nodes gets infected. This process is repeated for $10^3$ initial conditions.
In Figure~\ref{fig:SI_plots}, we plot the average fraction of infected nodes 
$\langle I(t)\rangle /N=\langle I(t;i_0,t_0)\rangle /N$ as a function of the time $t$ from $t_0$,
where the average is taken over the initial conditions $i_0$ and initial
infection times $t_0$. This quantity can also be interpreted as the probability
of nodes becoming infected at times shorter than $t$, $\langle |\{i|t_{\mathrm{inf},i} \le t \}| /N \rangle$. 
The inset of Fig.~\ref{fig:SI_plots} displays
the distributions of the times to full prevalence, \emph{i.e.}~all nodes becoming
infected, $t_{\mathrm{final}}=\max_i t_{\mathrm{inf},i}$. As the spreading process
itself is deterministic, the variance of these distributions originates from 
the random initial conditions.

First, we focus on the three temporal reference models that remove temporal
correlations while leaving the static network intact (\emph{equal-weight link-sequence
shuffled}, \emph{time-shuffled}, and \emph{uniformly random times}). 
Fig.~\ref{fig:SI_plots} shows that the spreading dynamics for those
event sequences that contain bursts (the original and equal-weight link-sequence shuffled sequences) are slower than
than those for the reference models from which 
burstiness has been removed (sequences from the time-shuffled and uniformly random times models). This clearly indicates that burstiness of the event sequences slow down the spreading process significantly. 
Further, the dynamics for the original sequence
and equal-weight link-sequence shuffled model closely resemble each other,
as do the dynamics for the time-shuffled and uniformly random times models.
The former similarity means that cross-link correlations have only a small
influence on the speed of spreading for most of the duration of the process.
The latter resemblance indicates that the daily pattern has also only minor effect 
on the spreading speed. However, when looking at the 
distributions of the full prevalence times shown in the
inset of Fig.~\ref{fig:SI_plots}, $t_{\mathrm{final}}$, 
it is seen that for long times, cross-link correlations speed up 
the process
somewhat. The full prevalence
times for the uniformly random times and the time-shuffled reference
models are almost indistinguishable. 

Next, we turn to those reference models 
that modify structural features of 
the static aggregated network. We find that when the network topology is
retained but weight-topology correlations are removed with the link-sequence
shuffled reference model, the spreading significantly speeds up compared to the original. 
This is because the reference model removes the known
Granovetterian weight-topology correlations where weak links
connecting dense communities of nodes act as bottlenecks 
~\cite{Onnela07}. In addition, if topological correlations such as the community structure
and geographical embedding are removed with the configuration model, the dynamics of 
spreading become even faster with the exception of the early stage of the process.
If we further remove the degree
heterogeneity which is retained in the configuration model by constructing 
an Erd\H{o}s-Renyi (ER) random networks (the random network model), there is again a small
increase when compared to the configuration model. However, the configuration
model is more efficient in spreading the infection than E-R networks
once the critical mass of infected nodes is reached, and when considering
only the full prevalence times, the configuration model is the fastest lattice for spreading.

Finally, we cross-compare the relative importance of the structural and temporal correlations
in the call sequence on the spreading speed. As seen in Fig.~\ref{fig:SI_plots},
the spreading dynamics for the time-shuffled model where weight-topology correlations
are retained but the bursts that are 
destroyed are faster compared to the link-sequence
shuffled model, where bursts are retained but weight-topology correlations are destroyed.
This means that the burstiness of call sequences on individual links
 plays a more important role than weight-topology correlations in slowing
 down the spreading dynamics. Furthermore, the results for the configuration
 model and the uniform times model are roughly comparable; the first one removes 
 all structural correlations while retaining all temporal correlations and the second one 
 removes all temporal correlations while retaining all structural correlations.
This observation can be considered surprising since the structural features of static
networks are typically considered to be the most important factor for any dynamical
process, while temporal correlations are assumed to have minor role.
However, the above points out that the temporal correlations may be 
of equal importance.

In addition, we have thoroughly investigated the robustness of the above results
by studying the role of the periodic boundary conditions, the initial conditions,
the system size, and the error of the mean (see \ref{sec:boundaryconditions} and 
\ref{sec:citynetworks}); no significant changes to the above results were observed.

\section{Spreading speed on single links}

In Section~\ref{sec:spreading}, we have shown that the 
temporal inhomogeneities of our empirical sequence slow down spreading
on the network and that considerable amount of the slowing down can be attributed
to the burstiness of event sequences of links. 
In order to better understand
the effects of temporal inhomogeneities, we will next focus on the event sequences corresponding to 
each of the links, and study the effects of temporal inhomogeneities
on the spreading speed at the link level. In order to do this,
we investigate the statistics of the \emph{relay times} of links.
The relay time of an edge denotes the time it takes for a newly infected node to spread the 
infection further via the next event that the link participates in.

\begin{figure}[t]
\includegraphics[width=1.0\linewidth]{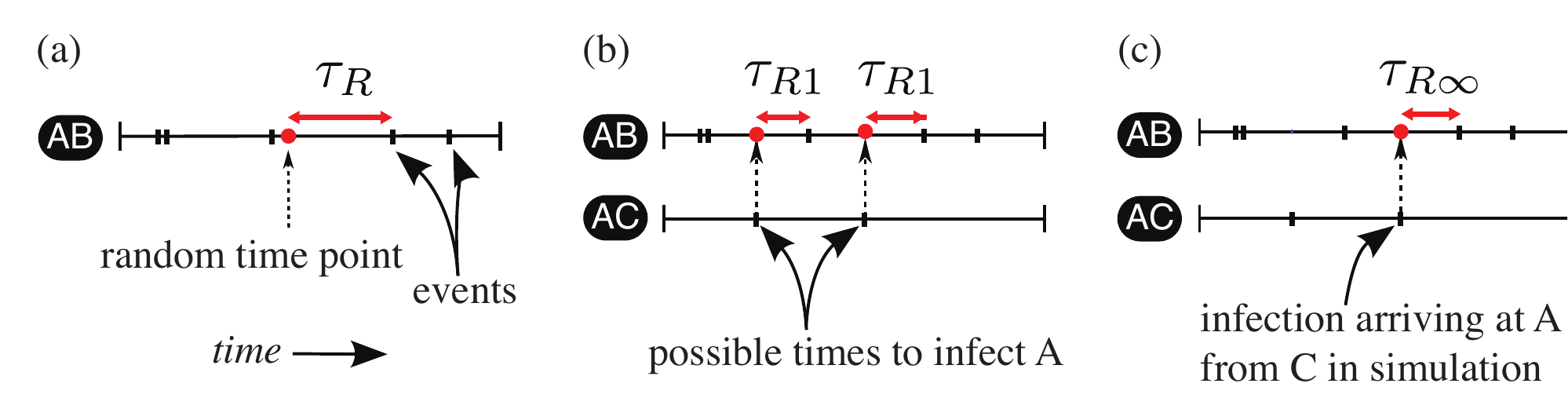}
\caption{Schematic illustrations of the three relay times.
The horizontal line represents a link, and the small
vertical lines displays the events between the two nodes within the
observation period, $t \in [0,T]$. (a) The random relay time $\tauzero$ is the time
difference between a randomly selected point in time and the time of the next
event. (b) The triggered relay time $\tauone$ for the link $AB$ is the time difference
between the time of an event on any other link of the two nodes (A or B) and 
 the time of the next event on the edge $AB$. (c) The actualized
 relay time $\tauinf$ is calculated on the basis of simulations
 and measures the time between the node becoming infected
 and further transmitting the infection.
}
\label{fig:tau_illustrations}
\end{figure}

In the first approximation, we will neglect the effects of correlated event sequences
on adjacent links and assume that nodes become infected at uniformly random times.
This approximation is captured by the \emph{random} relay 
time $\tauzero$, the waiting time between a uniformly random point in time and the 
next event on a link. 
Going beyond this approximation, nearest-neighbor correlations can be taken into account
via the \emph{triggered} relay time $\tauone$, where nodes may only get infected by the 
events they participate in. 
The triggered relay time is thus defined as the time
between a uniformly randomly picked event involving one of the two nodes of the focal link
and the next event on that link.
Finally, one can consider the effects of network-wide correlations and temporal
paths of the network, and measure from empirical data the actual 
times between the infection arriving at a node and crossing a link
in simulations, leading to \emph{actualized} relay times $\tauinf$. These
three concepts of relay times are schematically illustrated in Fig.~\ref{fig:tau_illustrations}.

\subsection{Random relay time $\tauzero$} 
\subsubsection{Theory}
We begin our analysis of relay times 
by neglecting all correlations between the event sequences of
adjacent links, and study the dependence
of the relay times on the inter-event time
distributions of the links. We consider the case where one
of the nodes, $A$ or $B$, of the link $AB$ becomes infected at a uniformly
random point in time. The random relay time $\tauzero$ is then defined as the
random variable representing the time between this 
random time of infection and
the time of the next event between $A$ and $B$. 
In this approximation, the relay times $\tauzero$ of a link are
determined purely by the inter-event time distribution of events on that link,
$P(\iet)$, where the inter-event time $\iet$ is defined as the time difference
between two consecutive events. On the basis of the inter-event time
distribution, the probability distribution of the relay times can be written as
$P(\tauzero)=\frac{1}{\avg{\iet}}
\int_{\tauzero}^{\infty}P(\iet) d\iet$ \cite{Vazquez07}. Further, 
assuming that the tail of the
distribution is not too broad ($\lim_{x \to \infty} x^2 P(\iet > x) = 0$),
the average relay time $\avg{\tauzero}$ can be calculated analytically for 
any inter-event time distribution:

\begin{equation}
\avg{\tauzero} =\int_0^\infty \tauzero P(\tauzero) d\tauzero=\frac{1}{2} \frac{\avg{\iet^2}}{\avg{\iet}}.
\label{t0_analytical}
\end{equation}
The value of $\avg{\tauzero}$ depends heavily on the average inter-event
time $\avg{\iet}$, or equivalently, on the average rate of events taking
place on the  link. As we are primarily interested in the effect of
the shape of the inter-event time distribution on 
$\avg{\tauzero}$, it is natural to use a Poissonian distribution as a reference. Thus, we define the normalized 
average relay time, $\avg{\tauzero}^*$, by dividing $\avg{\tauzero}$ by the corresponding
 relay
time for a Poisson process $\avg{\tauzero[,p]}$, for which the average
inter-event time is matched to that of the given inter-event time
distribution such that $\avg{\iet_p}=\avg{\iet}$. The normalized relay time $\avg{\tauzero}^*$ can then be written as 
\begin{equation}
\avg{\tauzero}^* \equiv \frac{\avg{\tauzero}}{\avg{\tauzero[,p]}}=\frac{\avg{\iet^2}}{\avg{\iet^2_p}}=\frac{\avg{\iet^2}}{2\avg{\iet}^2},
\label{r_definition}
\end{equation}
where $\avg{ \tauzero[,p]} =\avg{\iet_p}=\avg{\iet}$. 
Thus, $\avg{\tauzero}^*$ 
measures the ratio of the second moment to
the square of the first moment of the inter-event time distribution.
Generally, the broader the distribution, the larger the second moment
as compared to the square of the first moment. Because of this, $\avg{\tauzero}^*$ can
be viewed as a measure of 
 the broadness of the inter-event time distribution. Since a
broad inter-event time distribution is also a signature of a bursty time
sequence,  $\avg{\tauzero}^*$ can equally well be seen as a measure
of burstiness.  This is further illustrated by the following: $\avg{\tauzero}^*$ can be written
 as $\avg{\tauzero}^*=[(\frac{\sigma_{\iet}}{\avg{\iet}})^2+1]/2$, where $\sigma_{\iet}$
is the standard deviation and  $\frac{\sigma_{\iet}}{\avg{\iet}}$ the
coefficient of variation of the inter-event time distribution. 
The coefficient of variation $\frac{\sigma_{\iet}}{\avg{\iet}}$ was also used by Goh and Barab\'{a}si
\cite{Goh08} as a basis for their measure of burstiness, $B=(\frac{\sigma_{\iet}}{\avg{\iet}}-1)/(\frac{\sigma_{\iet}}{\avg{\iet}}+1)$.
Thus, the normalized average relay time can be written as
 $\avg{\tauzero}^*=\frac{B^2+1}{(B-1)^2}$. Hence, the burstier an event sequence is,
 the longer the random relay times are. 
 
 \begin{table}
\begin{center}
\begin{tabular}{|c|c|c|c|}
  \hline
$\sim p(\iet)$ & $\avg{ \tauzero }$ & $\avg{\tauzero}^*$ & Definitions\\
\hline\hline
$e^{-\iet / t_c}$ & $t_c$ & 1 & $t_c > 0$ \\
\hline
$\iet^{-\alpha},1<\iet$ & $\frac{1}{2} \frac{\alpha -2}{\alpha -3}$ & $\frac{(\alpha -2)^2}{2(\alpha -1)(\alpha -3)}$& $\alpha >3$\\
\hline
$\iet^{-\alpha},1<\iet< t_c$ & $\frac{\gamma(3)}{2\gamma(2)}$ &$\frac{\gamma(1)\gamma(3)}{2\gamma(2)^2}$ &$\gamma(k)=\frac{1-t_c^{k-\alpha}}{\alpha -k},\alpha \not= k$ \\
\hline
$\iet^{-\alpha}e^{-\iet / t_c},1<\iet$ & $\frac{t_c}{2} \frac{\theta(3)}{\theta(2)}$ & $\frac{\theta(1)\theta(3)}{2\theta(2)^2}$&\scriptsize{$\theta(k)=\Gamma(k-\alpha,t_c^{-1})$}\\
\hline
\end{tabular}
\end{center}
\caption{Functional forms of the random relay times $\avg{\tauzero}$ and
the normalized random relay time $\avg{\tauzero}^*$ for some inter-event time
distributions.}
\label{tau0_analytical}
\end{table}
 
 \begin{figure}[t!]
\subfigure{\includegraphics[width=0.46\linewidth]{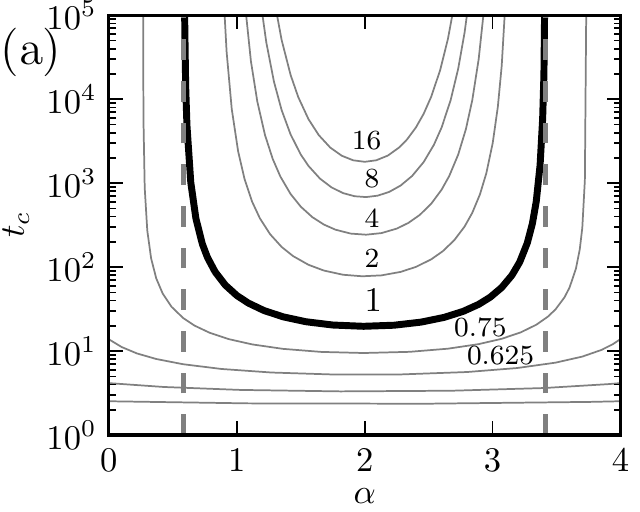}}
\subfigure{\includegraphics[width=0.46\linewidth]{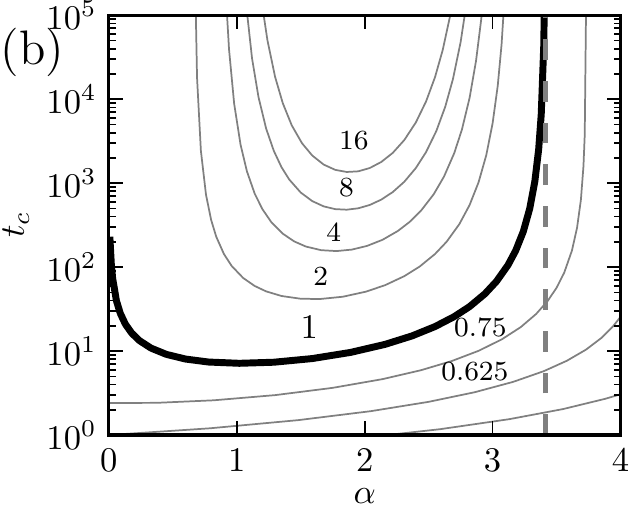}}
\caption{Contour plots the value of $\avg{\tauzero}^*$ when the inter-event time distribution is (a) 
power-law with exact cutoff and (b) power-law with exponential 
cutoff (See the Table~\ref{tau0_analytical}). 
The black solid contour lines correspond to values of
$\avg{\tauzero}^*=1$, the gray $n$'th line above that to $\avg{\tauzero}^*=2^n$ and the gray $n$'th line 
below the black line to $\avg{\tauzero}^*=\frac{1}{2}(1-2^{-n})$. The gray dotted vertical lines are at $\alpha=2\pm \sqrt{2}$. }
\label{fig:TauZeroAnalytical}
\end{figure}

The inter-event time distributions of empirical 
event sequences are often broader than would be expected from a Poisson process. Such empirical
distributions have been fitted with power laws~\cite{Wu10,Min11} or
power laws with an exponential cutoff~\cite{Vazquez07,Candia08}. In Table
\ref{tau0_analytical} we present analytical formulas for $\avg{\tauzero}$
and $\avg{\tauzero}^*$ for these distributions and illustrate their dependence
on the distribution parameters.
For the power-law distribution, $\avg{\tauzero}^*$ becomes infinite 
when the power-law exponent $\alpha=3$, and decreases with increasing $\alpha>3$,
reaching $\avg{\tauzero}^*=1$ when $\alpha=2+\sqrt{2}$. Thus, for power-law
inter-event time distributions in this regime, the average random relay times are
longer than for the Poissonian reference case. 
For power-law
distributions with cutoff, the cutoff value $t_c$ also affects
$\avg{\tauzero}^*$. The joint effect of the power-law exponent
and the cutoff is illustrated with contour plots in 
Fig.~\ref{fig:TauZeroAnalytical}. If
the cutoff point $t_c$ is small enough, the random relay times are
shorter than Poissonian also for power-laws with $\alpha \le 2+\sqrt{2}$, but for large values of $t_c$, the power-law
inter-event times dominate and the relay times become longer.

\subsubsection{Relay times in empirical event sequences}

Next, we turn our attention from the general case to empirical
data, where we have a finite number of events, $n$, in a finite observation
period, $t \in [0,T]$. For simplicity, we consider the case with periodic
temporal boundary conditions, where all events get repeated with a periodicity of
$T$. Then, the expression of relay time distribution of
Eq.~(\ref{t0_analytical}) becomes
$P(\tauzero)=\frac{1}{T}\sum_{i=1}^n[\tauzero \le \iet_i]$, where
$\iet_i$ is the inter-event time between the $i-1$-th and $i$-th events.
The expected relay time $\avg{\tauzero}$ is then given by
\begin{equation}
\avg{\tauzero} =\frac{\sum_{i=1}^n\iet_i^2}{2T}.
\label{tau0_iee}
\end{equation}
For a given values of $n$ and $T$, the average relay time $\avg{\tauzero}$ is
minimized when all the inter-event times are the same, that is, $\iet_i=\iet_j,
\forall i,j$, giving $\avg{\tauzero} =\frac{T}{2n}$. In contrast, it is
maximized when all the events happen at the same time resulting in a single
non-zero inter-event time, that is, $t_i=t_j \forall i,j$ and $\iet_1 = T$
giving $\avg{ \tauzero } = T/2$. Thus, $\avg{\tauzero}$ is maximized when
all events take place in the same burst and $\avg{\tauzero}$ is minimized
when events are maximally separated in time.

As with the general case, we want to compare the $\avg{\tauzero}$ values to some baseline. Here, the proper baseline is to distribute the times of the $n$ events of each edge at uniformly random within the observation period, equivalently to the uniform random times reference model defined earlier. This is equivalent to having a Poisson process with a constant average inter-event time with the additional condition that there are $n$ events in the interval \cite{Steutel67}. Order statistics can be used to analytically derive the inter-event time distribution, which in this case is a beta distribution \cite{David03} (see \ref{sec:order_statistics} for details). The beta distribution converges to an exponential distribution as the number of events $n$ increases, and thus the uniformly random times reference model approaches the Poisson process with increasing $n$.  The expected value and variance of $\avg{\tauzero}$ for the uniform times model with periodic boundary conditions are given by
\begin{equation}
\ensavg{\avg{\tauzero}} = \frac{T}{n+1}, \operatorname{Var}(\avg{\tauzero})=\frac{T^2(n-1)}{(n+1)^2(n+2)(n+3)},
\label{eq:uniftimes_tauzero}
\end{equation}
and the normalized relay time (similarly to Eq.~\ref{r_definition}) becomes
\begin{equation}
\avg{\tauzero}^*=\frac{\avg{\tauzero}}{\ensavg{\avg{\tauzero}}}=\frac{(n+1)}{2}\sum_{i=1}^n(\iet_i/T)^2.
\label{eq:normalized_tauzero}
\end{equation}
Further, since the normalization of $\avg{\tauzero}^*$ is done with the uniformly random times reference model instead of a Poisson process, we need to include a correction factor in the relationship between burstiness and the normalized relay time: $\avg{\tauzero}^*=\frac{n+1}{n}\frac{B^2+1}{(B-1)^2}$. 

\begin{figure}[t!]
\subfigure{\includegraphics[width=0.499\linewidth]{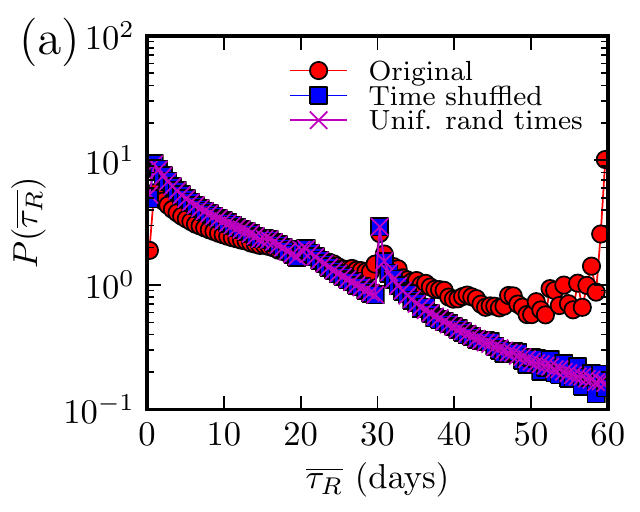}}
\subfigure{\includegraphics[width=0.499\linewidth]{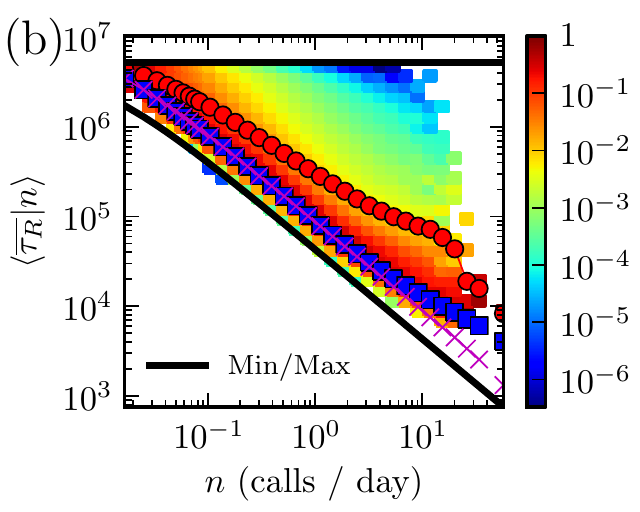}}
\subfigure{\includegraphics[width=0.499\linewidth]{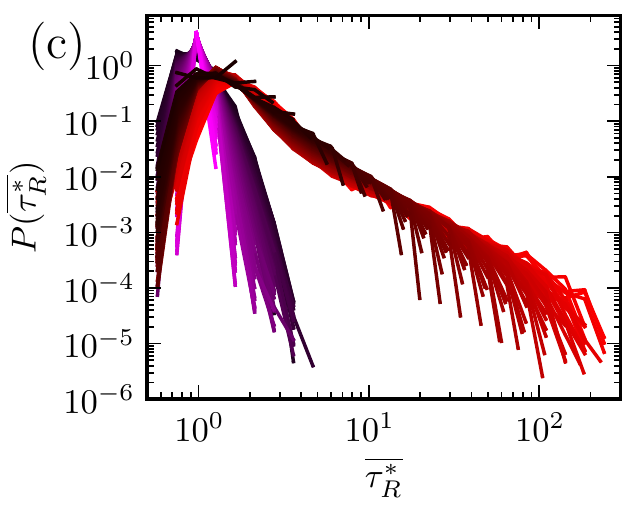}}
\subfigure{\includegraphics[width=0.499\linewidth]{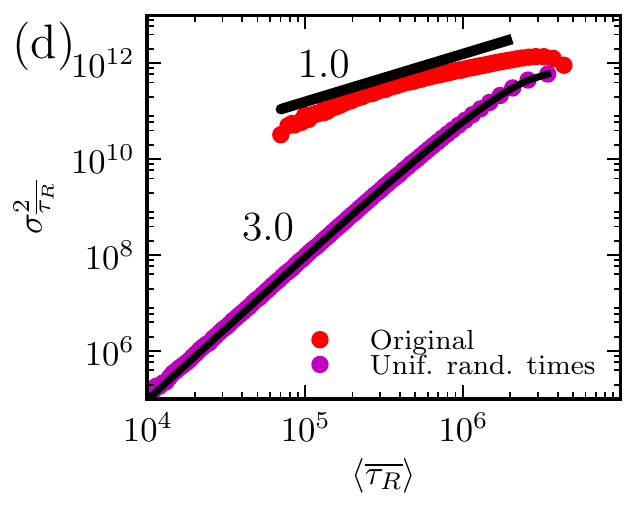}}
\caption{(a): Distribution of the relay times $\avg{\tauzero}$ for
the original, time-shuffled and uniformly random event sequences. The peak
at $\sim 30$ days is due to the larger amount of edges with only 2 calls (a
result of the mutuality condition) and periodic temporal boundary
conditions. (b): $\langle \avg{\tauzero} | n\rangle$ for
the original event sequence and reference models.
For the original sequence, we also show the probability distributions
of $\avg{\tauzero}$ conditional to weight (see color bar on right). The solid
lines denote (theoretical) max and min values. 
(c): The distributions of $\avg{\tauzero^*}$ conditional to (logarithmically binned) weights for the original sequence
(red) and the uniformly random times model (magenta). Each curve is
colored according to the weight associated to it such that that dark colors
indicate small weights. 
(d): The average and the variance of $\avg{\tauzero}$ for each of the bins.
The solid line on the top indicates 
power law fit to the original data, and the lower black line corresponds to the
analytical values for the uniformly random time sequence calculate with Eq.~\ref{eq:uniftimes_tauzero}.
Distributions for bins with more than 800 calls during the time interval $T$ are not shown due to insufficient statistics.}
\label{fig:numCallsVsTauZero}
\end{figure}

Next we calculate $\avg{\tauzero}$ for each edge for the empirical MPC sequence using Eq.~(\ref{tau0_iee}) and compare the values to the temporal reference models, where the static network remains unchanged. Further, we leave out the equal-weight link-sequence shuffled model which would give equivalent results to the empirical sequence. This leaves us only with the time-shuffled and the uniformly random times reference models. In Fig.~\ref{fig:numCallsVsTauZero}~(a) we plot the probability distribution of $P(\avg{\tauzero})$ for the original data and for the random references. It is seen that in the empirical data there are more edges with high random relay times and fewer edges with low relay times than in the reference models.
This is also reflected in the value of the network-level average of $\avg{\tauzero}$ for the original sequence, averaged over all edges: it is 
 $23.1$ days while for the reference models the average value is only $13.8$ days.  Thus, the high average relay times of the empirical network can at least partly explain the slowing down of the SI-process. 
 Further, the distributions of $\avg{\tauzero}$ for the time-shuffled and uniformly random model are indistinguishable, as expected.

The random relay times of edges depend heavily on their weight $n$, \emph{i.e.}~the total number of calls on each edge. 
Thus the shapes of the $\tauzero$ and $\avg{\tauzero}$ distributions are dominated by the heterogeneous weight distribution.  In Figure~\ref{fig:numCallsVsTauZero}~(b) we plot $\avg{\tauzero}$ as a function of the edge weight. 
The difference between the original event sequence and the reference sequence grows with the edge weight, and the edges on the original sequence are on average slower than on the references independent of the weight.
However, for links of the same weight, the  $\avg{\tauzero}$ values for the empirical data are broadly distributed, and the tail of this distribution is partly responsible for the longer average relay times.
Note that for the time-shuffled reference model, the average relay time values are almost the same as for the uniform random times model except
for very high weights, where the time-shuffled reference is slower. 
As other temporal correlations than the daily pattern are absent from this reference, any deviation is due to the daily pattern. It is seen that the daily pattern starts to have an effect only when the call frequency is larger than a few calls per day. More than 99\% of the edges in the network have less than three calls per day, and thus for a vast majority of the edges, the daily pattern has practically no effect on the relay times, in line with our observations on spreading dynamics.

As seen in Figure~\ref{fig:numCallsVsTauZero}~(b), the random relay times display a lot of variation even for edges of the same weight. We illustrate this variation in more detail in Figure~\ref{fig:numCallsVsTauZero}~(c), and show the distributions of $\avg{\tauzero^*}$ for edges with fixed weight. For high-weight edges, we use a weight range to improve statistics. Remarkably, the distributions for the original event sequence collapse together for all weights, with the exception of a small offset in the peak and differences in cutoff values ($\mathrm{max}(\avg{\tauzero^*})=(n+1)/2$). Thus there is a signature distribution of the relay times that does not depend on edge weight -- as seen in comparison with the uniformly random times reference model, this distribution is broad and the relay times contain a lot of variation. Thus, while many edges have relay times close to the reference model, there is a small number of highly bursty edges with very long relay times. 
Furthermore, the length of the tail of the $\avg{\tauzero^*}$ distributions grows with the edge weight for the original sequence, which is in contrast to the uniformly random times sequence for which the variance is smaller for higher weights. To better quantify the difference in relay times within the group of edges with equal weight in Figure~\ref{fig:numCallsVsTauZero}~(d) we show a scatter plot where each point corresponds to the variance and the average of $\avg{\tauzero}$ for an edge group (similar to fluctuation scaling \cite{Eisler08}). For the uniformly random times we can use Eq.~(\ref{eq:uniftimes_tauzero}) to calculate the relationship between the average and the variance. For large weights we get $\ensavg{\avg{\tauzero}} \propto (\sigma^2_{ \avg{\tauzero} } )^\alpha$, with $\alpha=3$. However, for the original sequence the scaling exponent is close to $\alpha=1.0$ and the variance is higher than for the reference model for all values of $\ensavg{\avg{\tauzero}}$.

\subsection{Triggered relay time $\tauone$}
\subsubsection{Theory}

For the random relay times $\tauzero$ considered in the previous 
section, it was assumed that the event sequences on adjacent links
are uncorrelated, and thus the infection may arrive at one of the nodes
of the focal link at a random point in time. We will next take into account the  effects of
correlations between the times of events of adjacent links,
and require that a node may only become infected when it interacts
with one of its neighbors. Instead of choosing the time of infection of node $A$ or $B$ of the link $AB$
randomly from the time interval $[0,T]$, we pick it randomly from the set of
times where $A$ or $B$ participates in an interaction event with one of its neighbors $C \neq B$ and $C \neq A$ \cite{Miritello10}.
Then, the \emph{triggered} relay time $\tauone$ for the link $AB$ is defined as the time
difference between the infection arriving $A$ (or B) to the time when the infection is
passed through the link to $B$ (or A), similarly to Ref.~\cite{Miritello10}.

The average triggered relay time $\avg{\tauone}$ for the link $AB$ depends on both the 
time intervals between the events on the link, as well as correlations of
the times of events between the link and its neighbors.
 If the events on the other links are independent of the events on $AB$, 
 then $\avg{\tauone}$
approaches $\avg{\tauzero}$. 
We illustrate the differences between  $\tauone$ and
$\tauzero$ with the help of a simple model that involves correlations 
between the sequences. 
We define a \emph{triggering model} where the system consists of two
links, AB and AC, such that each of the two links has $N$ events. We
assign times to the events on both links in a pairwise fashion, such that
each AB-AC event pair is considered to be triggered with a probability $p$ and uncorrelated
with probability $(1-p)$. For uncorrelated event pairs, we assign times
drawn uniformly at random from the interval $[0,T]$. For correlated event pairs,
we first choose the time $t_{AB}$ of the event on the AB link uniformly at random.
Then we set an event on the AC link randomly at either $t=t_{AB}-\delta t$
or $t=t_{AB}+\delta t$,  i.e. the AC event is considered to trigger the AB event or vice versa.
The time $\delta t$ is picked at random from some triggering
time distribution.
Periodic temporal boundary conditions are invoked if the AC event falls outside the interval $[0,T]$.
Thus, on both links we have on average $pN$ events that are triggered or trigger another event,
and $(1-p)N$ uncorrelated events.

Now we can calculate for the expected triggered relay time $\ensavg{\avg{\tauone}}$ for the AB link
by assuming that the trigger times $\delta t$ are small compared to the random
inter-event times, \emph{i.e.}~we always have $\iet > \delta t$. Note that the model is symmetric in 
that $\ensavg{\avg{\tauone}}$ is same for both links, AB and AC.
The fraction of uncorrelated events is $(1-p)$, and for these, 
the expected triggered relay time is simply given by Eq.~(\ref{eq:uniftimes_tauzero}): $\ensavg{\avg{\tauone}}=T/\left(n+1\right)$. 
The fraction of
triggered event pairs is $p$; half of the events are triggered by an AC event with a triggered
relay time $\ensavg{\delta t}$. The rest trigger AC events, 
and the expected time from the triggered AC event to the next AB event is $\ensavg{\iet-\delta t}=\frac{T}{n}-\ensavg{\delta t}$.
Thus, considering all these cases, the expected triggered relay time 
\begin{eqnarray}
\ensavg{\avg{\tauone}} &= & (1-p) \frac{T}{n+1} +\frac{1}{2}p \ensavg{\delta t} + \frac{1}{2}p (\frac{T}{n}-\ensavg{\delta t}) \nonumber\\
& = & (1-\frac{p}{2}\frac{n-1}{n})\ensavg{\avg{\tauzero}}.
\label{tau1_model}
\end{eqnarray}
Thus, the more triggered events there are, the shorter the triggered 
relay times are on average, as expected. Note that in the above equation $\ensavg{\avg{ \tauone}}$ does not depend on the
distribution of $\delta t$ as long as the values of $\delta t$ are small enough
compared to the inter-event times. 

\begin{figure}[t!]
\subfigure{\includegraphics[width=0.499\linewidth]{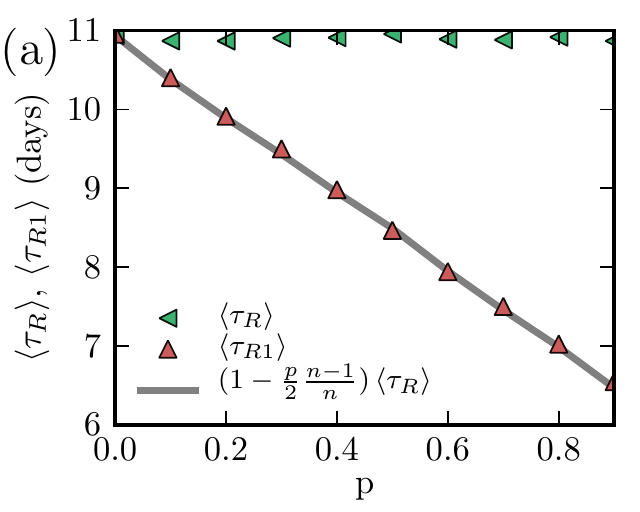}} 
\subfigure{\includegraphics[width=0.499\linewidth]{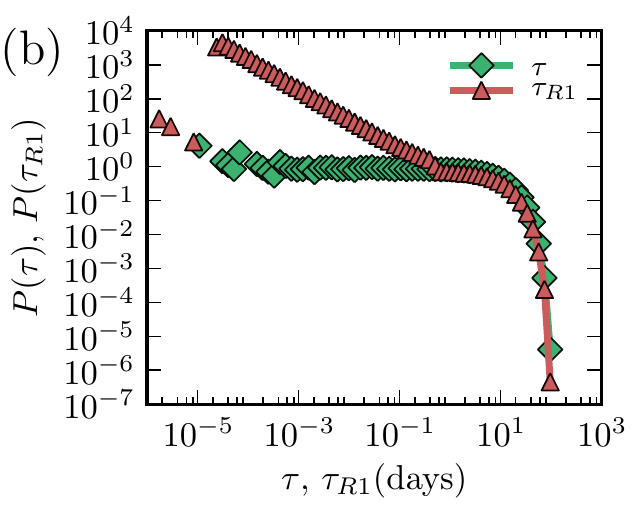}}
\subfigure{\includegraphics[width=0.499\linewidth]{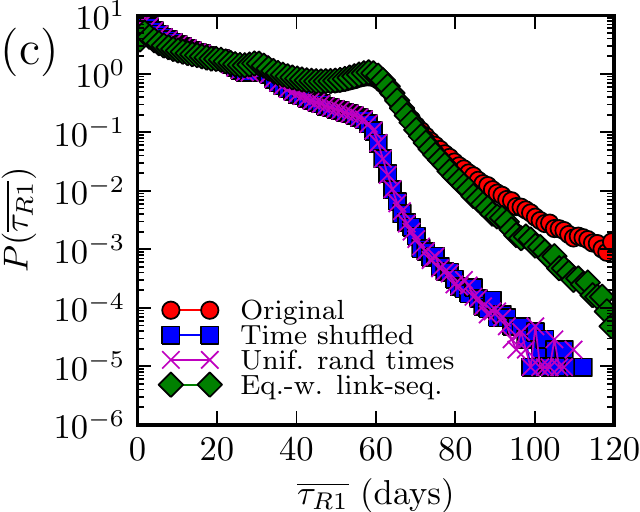}}
\subfigure{\includegraphics[width=0.499\linewidth]{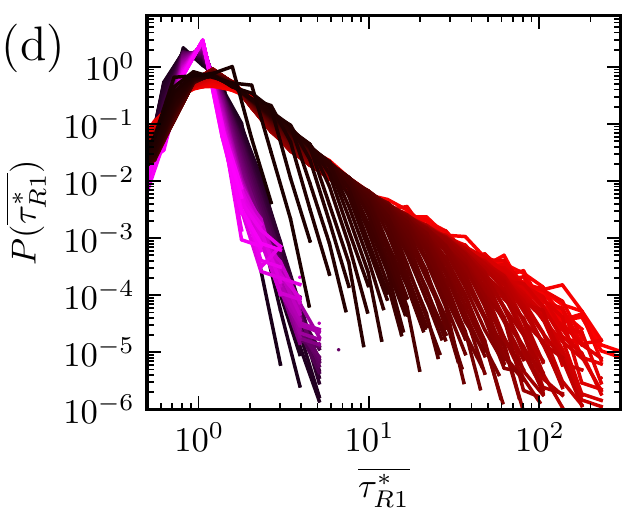}}
\caption{(a)-(b) Results of the triggering model with parameters $T=120$ days,
$N=10$, while $P(\delta t)$ follows a power-law with exponent $0.9$ and
cutoff at $10^5$ seconds. The simulations are performed over $10^4$
independent realizations. (a) The distribution of $\langle\avg{\tauzero}
\rangle$, and $\langle\avg{\tauone}\rangle$ as a function of the coupling
parameter $p$. The solid curve represents 
Eq.~(\ref{tau1_model}). (b) The distributions of $\tauone $ and inter-event
time $\iet$ for the coupling parameter $p=0.4$.  Inter-event times are
exponentially distributed wheres the distribution of $\tauone $ follows a
power-law with exponential tail.
(c)Distributions of the average relay time in empirical data, $P(\avg{\tauone})$, for the original event sequence and the temporal reference models. 
(d) The distributions of $\avg{\tauone^*}$ conditional to (logarithmically binned) weights for the original sequence
(red, curves on the right) and the uniformly random times model (magenta, curves on the left). Each curve is
colored according to the weight associated to it such that that dark colors
indicate small weights.
}
\label{couplinmodel_results}
\end{figure}

The behavior of the triggering model is illustrated in Fig.~\ref{couplinmodel_results}~(a),
where the average random and triggered relay times from simulations of the model
are shown; the latter are seen to match well with the 
analytical prediction of Eq.~(\ref{tau1_model}). For these simulations, we have 
used a power-law distribution of the triggering times $\delta t$ with an exponent
of 0.9 and a cutoff at $t_c=10^5$ seconds; this has been motivated by the shape
of the low-$\Delta t$-region of triggered relay time distributions in experimental 
data (see, e.g.~\cite{Miritello10}). 
Fig.~\ref{couplinmodel_results}~(b) shows the distribution of the
inter-event times $P(\iet)$ and the relay times $P(\tauone)$ for 
the model with this
triggering time distribution. The distribution
$P(\tauone)$ shows an initial power-law decay which is caused by the
triggered events, followed by exponential decay. In contrast, the
inter-event time distribution $P(\iet)$ is exponential.

\subsubsection{Empirical results}

Next we study the averages of triggered relay times $\avg{\tauone}$ in empirical data.
In Fig.~\ref{couplinmodel_results}~(c) we plot the distributions of $\avg{\tauone}$ of the
MPC network original call sequence and the temporal reference models. Similar to random relay times,
the distributions of $\avg{\tauone}$ for the time-shuffled and uniform random time model match with each other. Further,
the distribution for the original event sequence has a fatter tail than these two reference models.
Unlike with the case of $\avg{\tauzero}$, the distributions of $\avg{\tauone}$ are not same for the original and the equal-weight
link-sequence shuffled event sequences. This is because equal-weight link-sequence shuffling destroys the correlations between events on neighboring edges.
In the coupling model, such correlations caused the expected $\avg{\tauone}$ values to decrease.
This is true also for the MPC network, since the average $\avg{\tauone}$ for the original sequence is $22$ days, where it is
$23$ days for the equal-weight link-sequence shuffled event sequence.

In Fig.~\ref{couplinmodel_results}~(d) we plot the $\avg{\tauone^*}$ distributions conditional to the edge weight similar to
Fig.~\ref{fig:numCallsVsTauZero}~(c). The main difference to $\avg{\tauzero^*}$ distributions is that the theoretical maximum value
for $\avg{\tauone^*}$ is $n+1$ instead of $(n+1)/2$. The values of  $\avg{\tauone^*}$ inside the interval from $(n+1)/2$ to $n+1$ 
arise from situations where the triggering events take place close to the beginning of a long interval between consecutive bursts.
Thus, they result from a combined effect of burstiness and timings of triggering events. 
Such situations are not very frequent, leading to steep decays in the tails of the $\avg{\tau^*_{R1}}$ distributions for values larger than $(n+1)/2$.
Thus, even for small edge weights, the tails of the  $\avg{\tau^*_{R1}}$ distributions do not follow the signature scaling observed for  $\avg{\tau^*_{R}}$ distributions.

\subsection{Actualized relay time $\tauinf$}
\subsubsection{Motivation: the event infectivity $\eventinf$}
When defining the triggered relay time $\tauone$, it was
assumed that each event taking place on the neighboring edges
transmits the infection to the focal edge with an equal probability.  In this section, we
study the validity of this assumption by defining a quantity called
\emph{infectivity}, $\eventinf$, for each event. The event infectivity
is defined as a probability of an event spreading the infection
during a spreading process which is initiated from a random node and a random point in time.

In order to calculate exact values for the event infectivities,
we would need to simulate the spreading process over the entire ensemble of
all possible initial conditions. Instead, we calculate an estimate for
$\eventinf$ by starting the infection from $10^4$ randomly selected initial
conditions and simulating each spreading process until all the nodes in the
network are infected. For these processes, we calculate the number of
times each event transfers the infection to an susceptible node. 
An estimate for the event infectivity, $\hat{\eventinf}$, is then determined by
dividing these numbers by the number of initial conditions. 
Note that  this approximation is not very accurate for small values of $\eventinf$.

\begin{figure}
\subfigure{\includegraphics[width=0.332\linewidth]{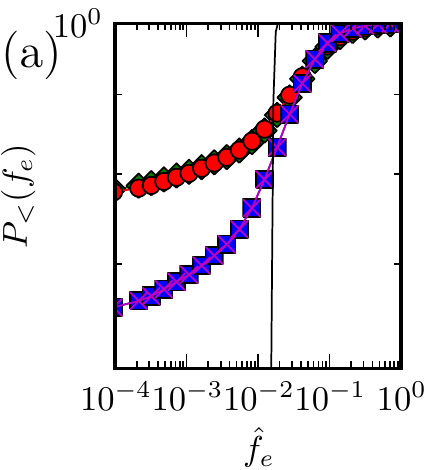}}
\subfigure{\includegraphics[width=0.332\linewidth]{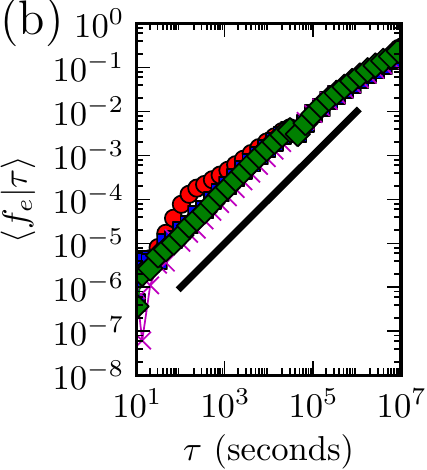}}
\subfigure{\includegraphics[width=0.332\linewidth]{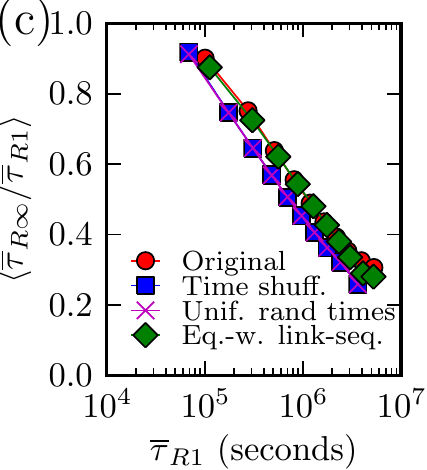}}
\caption{(a) The cumulative distribution of estimated event infectivity, $P_<(\hat{\eventinf})$, for the
original sequence and the temporal reference models. The black line corresponds to the mean of the estimated infectivity distribution
for a constant "real" event infectivity; the variation arising from sampling is small enough to render the standard deviation of this
distribution negligibly small. (b) The average event
infectivity as a function of the trailing inter-event time (time to last
event). The solid black line corresponds to a linear increase.
(c) The average ratio $\langle \avg{\tauinf} / \avg{\tauone}\rangle$ as a function of $\avg{\tauone}$. The edges are placed
into bins according to their $\avg{\tauone}$ values in a way that each bin contains equal number of edges. The coordinate of the bin
in the x-axis is then calculated as the average $\avg{\tauone}$ value of the edges inside the bin.
}
\label{infectivity}
\end{figure}

In Fig.~\ref{infectivity}~(a) we plot the cumulative distributions of the estimated event
infectivities $P_<(\hat{\eventinf})$ for the original sequence and the temporal reference models.
We find that there are large differences in the number of non-infecting events (events that have $\hat{\eventinf}=0$, \emph{i.e.}~were never observed to transmit the infection). The total fraction of events which do not
participate in the spreading process is $\sim 77$
\% for the original sequence and the equal-weight link-sequence shuffled model, while for
the time-shuffled and the uniform time model it is $65$\%. 
Further, if one assumes that all the ``real'' event infectivity values would be equal,  
the sampled event infectivity estimates would be distributed
with a mean $\langle \hat{\eventinf} \rangle =0.015$ and a standard deviation $\sigma(\hat{\eventinf})=10^{-3}$ 
\footnote{If there are $N$ nodes, $E$ events and the infection is  started with $L$ initial conditions, then for each event, the number of times it participates in the spreading, $C$, is binomially distributed with parameters $p=\frac{N-1}{E}$ and $n=L$. Then, for that event $\hat{\eventinf}=C/L$. Thus $\langle \hat{\eventinf} \rangle = p$ and $\sigma(\hat{\eventinf})=\sqrt{\frac{p(1-p)}{L}}$. The $\hat{\eventinf}$ values for the events are not independent of each other. However, the correlations between them are weak if $N \ll E$, and the $\hat{\eventinf}$ distribution gotten from the sampling process can be approximated by the distribution that a single event would have in this process.}. The observed variation
in the estimated infectivities is much larger than this. Thus we can conclude that the 
assumption of all events being equally probable carriers of infection is not valid -- there
is considerable variation that arises from the timings of the events.

To better understand which events are most or least likely to spread the infection, we plot in Fig.~\ref{infectivity}~(b) the average estimated event infectivity
as a function of the preceding inter-event time of the event, \emph{i.e.}~the time to the 
previous event. On average, the
infectivity of an event increases with the inter-event time. This means
that events participating in bursts (events with short preceding
inter-event times) are less likely to participate in the infection
spreading process as compared to events with long preceding inter-event
times. The increase in $\hat{\eventinf}$ with $\tau$ shows almost a linear relationship
for the temporal reference models; a linear relationship would be expected if the
infection times were random. This is simply because if we select a random
infection point in time, the probability $f_e$ of the infection going through
an event $e$ would be proportional to the preceding inter-event time of that
event $\iet_e$, that is $f_e\propto \iet_e/T$.  However, the curve for the original
event sequence shows a slight deviation from this linear dependency for short
inter-event times.

\subsubsection{Definition of $\tauinf$ and empirical results} 
The above observations on large variations in event infectivity and correlations between
preceding inter-event times and infectivities are not consistent with the assumptions behind
the definition of the triggered relay time $\tauone$, where it was assumed that the infection can be received through
each event with equal probability. Further, $\tauone$ doesn't take into account the possibility
of both nodes of the focal edge getting infected independently, without transmission through that edge,
as the spreading front expands through their neighboring nodes.

In order to account for the infectivity distributions and all other constraints during the spreading process we define 
the \emph{actualized relay time} $\tauinf$ as the time that the
infection actually waits on one of the nodes of an edge before crossing it in a network-wide spreading process, \emph{i.e.} the 
difference between the time when the infection actually arrives at one of the endpoint nodes of an edge
and the infection being transmitted through that link.
The values for $\tauinf$ are obtained numerically by running the spreading process
through the network, monitoring which nodes infect which other nodes, and 
calculating the relay times accordingly.
Note that only events transmitting the infection are considered for
$\tauinf$. As an example, if node $u$ is infected and the next event between $u$
and $v$ occurs after node $v$ has already become infected by some other node $w$,  
that event doesn't affect the $\tauinf$ value of the edge linking $u$ and $v$.

For estimating $\tauinf$ in empirical data, we use the same sampling scheme as previously used while estimating the infectivity values
$\eventinf$. In Fig.~\ref{infectivity}~(c) we plot the ratios $\ensavg{\avg{\tauinf} / \avg{\tauone}}$ as functions of the respective relay times, for the 
original sequence and the reference models.
 On average, the actualized relay times $\avg{\tauinf}$ are shorter than
the triggered relay times, and the ratio decreases with increasing $\avg{\tauone}$.
 Thus, when the network-wide spreading process is accounted for,
  the relay times speed up especially for edges that appear slow when
  measured with triggered relay times. This
  is because the pathways taken by the spreading process often avoid and circumvent the slowest edges.
However, in some cases it is not  possible to employ any faster shortcuts, \emph{e.g}~an edge leading to a 
node of degree one is always used in the spreading
process. Note that $\avg{\tauinf}$ is not defined for edges that only have zero infectivity events,
\emph{i.e.}~for edges that never transmit the infection.

\subsection{Comparing the relay times}

\begin{figure}[t]
\includegraphics[width=1.0\linewidth]{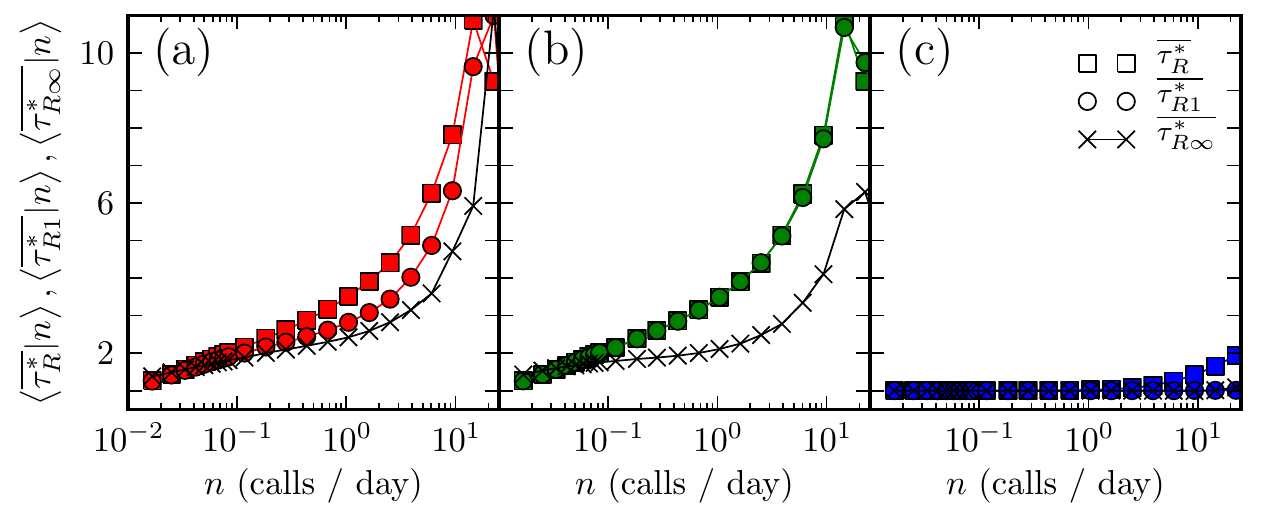}
\caption{The expected normalized average relay times $\langle \avg{\tauzero^*} |n \rangle$ (squares), 
$\langle \avg{\tauone^*} |n \rangle$ (circles), and  
$\langle \avg{\tauinf^*} |n \rangle$ (crosses) for the (a) original event sequence, (b) equal-weight link-sequence shuffled and (c) time-shuffled reference models. 
}
\label{fig:relativeTaus}
\end{figure}

Having introduced the three different relay times and illustrated their differences, we now turn to comparing them in detail with regards to the empirical MPC data. For the comparison, we normalize the average relay times with the expected times for the uniformly random times reference model.
The normalized random, triggered, and actualized relay times as a function of edge weight are shown in Figs.~\ref{fig:relativeTaus}~(a), (b) and (c) for the original event sequence, equal-weight link-sequence shuffled reference and time-shuffled reference, respectively.

Let us first focus on the scale of individual links and the behavior of the random relay times $\tauzero$ that do not take correlations between links into account. For the original sequence, the random relay times are always longer than for the uniformly random sequence, even for the smallest weights. Further, the difference grows with edge weight, reaching a maximum at $n \sim 12$ calls per day where the random relay times of the original sequence are on average $11$ times longer than for the reference. These effects reflect the burstiness of call sequences of edges. For the equal-weight link-sequence-shuffled reference, the random relay times are necessarily equal to the original sequence. For the time-shuffled reference, the random relay times deviate from the uniform case only for the highest-weight links. This means that the daily pattern has some effect on the highest-weight links, because it is the only difference between the two reference models; for the majority of links, it has no visible effect.

Next, we observe how the relay times change when information about events in the local neighborhood of the focal edge is taken into account with the triggered relay times $\tauone$. For the original event sequence, considering the timing correlations of neighboring edges decreases the relay times,
\emph{i.e.} speeds up the spreading process. This effect is negligible for small weights, but the difference to random relay times reaches $\sim$ 20 percent for edges with approximately one call per day. Note the difference between $\tauzero^*$ and $\tauone^*$ for the time-shuffled reference at this point is still negligible, and thus general timing correlations related to the daily pattern cannot be an explanation; rather, this difference is caused by calls triggering further calls.
For the equal-weight link-sequence shuffled reference,  the relay times $\avg{\tauzero^*}$ and $\avg{\tauone^*}$ appear the same. This is because correlations
between adjacent links have been destroyed; note that, however, in theory the network-wide daily pattern that is present in this null model could still induce
enough correlations to make a difference, but this is not the case. For the time-shuffled reference model, the slowing down effect of the daily pattern disappears completely for the triggered relay times. This happens because for $\tauone$, the triggering infection is more likely to arrive at the edge during daytime when also the focal edge is likely to be more active, and the inter-event times are thus shorter. On the contrary, $\tauzero$ increases because infections arriving at random points in time may have to wait long times during the nights before crossing the edge.

Finally, we consider the largest scale with the actualized relay times that take network-level spreading process and its timings into account.
For normalization, we cannot simply use Eq.~(\ref{eq:uniftimes_tauzero}) as the expected $\avg{\tauinf}$ values for the uniformly random times reference
model depend on the topology of the underlying network and other large-scale correlations in the call sequence. 
Instead, we use the $\tauinf$ values measured from 
simulations to define the normalized relay time as $\ensavg{\avg{\tauinf^*}}[n] = \ensavg{\avg{\tauinf}}[n] / \ensavg{\avg{\tau_{R\infty,unif. random}}}[n]$.
For both the original call sequence and the equal-weight link-sequence shuffled reference, the actualized relay times are shorter than random
or triggered relay times, except for the lowest-weight links. The difference grows with increasing weight. Thus, the slowing-down effect of bursts on the actualized relay times is smaller than would be expected just by looking at the links in isolation ($\tauzero$) or additionally the immediate neighborhood of the links ($\tauone$).
One possible explanation for the shorter actualized relay times on strong edges is the fact that the slower an edge is, the more often alternative spreading routes are employed making the progress of spreading faster. Because high-weight edges are much burstier than low-weight edges in relation to the uniform reference model, alternative routes are used more often for edges with high weights. 
Even though the normalized actualized relay times are smaller than the corresponding triggered relay times, the qualitative conclusion drawn from the triggered relay times still holds.
That is, because the call sequences on edges are bursty, the speed of SI spreading (or equivalently, the duration of fastest temporal paths) is exaggerated if no temporal information about the calls is considered but the calls are assumed to be distributed uniformly random within the observation interval. This is specially true for high-weight edges which slow down more than low-weight edges. Furthermore, the spreading process is not affected much by the daily pattern.

\section{Discussion}
In temporal networks, both the topology of the aggregated network and the timings of interaction events can be
crucial in determining how a 
dynamic process mediated by the network unfolds. 
We have explored the limiting case of the speed of spreading in the SI model, set up such that an event between an infectious and susceptible individual
always transmits the infection. In this process one of the nodes
is initially set to the infectious state, and the spread of infection follows the fastest time-respecting paths~\cite{Holme05, Pan11} from
this node to all other nodes. Thus the speed of the process sets an upper bound for the speed of any 
dynamic process that is mediated through the interaction events of the network.

The roles of different types of network correlations in the outcome of a dynamical
process can be studied with the help of reference models that switch off
selected correlations one by one. For static networks, structural correlations can be 
removed using the configuration model procedure, while weight-topology
correlations can be addressed by shuffling weights. The removal of 
time-domain correlations in temporal networks is still a fairly new problem, and although
several temporal reference models have been designed~\cite{Holme05,Karsai11, TempReview},
work in this respect can still be considered to be ``in progress''. In addition to the 
MPC case study, the contribution of this article in more general terms is to formalize
and make the use of temporal reference models systematic; we have also 
analytically calculated some key statistics for the uniformly random times
reference model.

The first observation made by using the framework of reference models
was that for the MPC network the timings of the sequence of 
events are as important as the topological structure of the network.
Randomizing either the topological structure or call sequences speeds up the
spreading process by 
roughly equal amount. Thus, when estimating the spreading velocity,
disregarding the time-domain information
and simply aggregating over it results in an error comparable
to the error that would result from
not taking the network topology into account. In general,
this observation points out the importance of taking time-domain
information into account when studying dynamics on temporal
networks;  e.g.
a similar slowing down has also been observed
for an email network, whereas temporal correlations give
rise to faster temporal paths in an air transport network
because of optimized scheduling \cite{Pan11} .

The second observation made with the reference models was that a large part of the slow dynamics on the MPC network can be
attributed to the call sequences of individual links. Thus, the slowing-down effect can
be explained by characterizing the timings of the call sequences of links.
For this, we have used the concept of relay times~\cite{Vazquez07,Miritello10}, and showed
how the burstiness and the slow spreading speed compared to the 
Poisson process are equivalent concepts: the slowing down is due to the bursty nature of the event sequences.
We also found that there is a lot of heterogeneity in the relay times of links even when normalizing by weight; thus
not only the broad distribution of link weights but also the distribution of relay times affects the speed of spreading.
Further, keeping the exact time sequence of calls on links, rather than considering them uniformly randomly distributed, slows down the high-weight links much more than links with low weights. This points out that the effect of the link weights to the spreading speed would be overestimated without information on event timings.

At the scale of individual links, the spreading speed through a link can be approximated with the random relay time
if we only consider information of the events on the focal link. 
This approximation can be improved by increasing the scale to include neighboring links by 
adding information about correlation between adjacent event sequences using
 the triggered relay time \cite{Miritello10}.
We have constructed a model that shows how such correlations decrease 
the average relay times; 
using triggered relay times also corrects for the artificial effect of the circadian and weekly patterns.
Finally, we have introduced the actualized relay time, motivated by empirically observed high variations in the likelihood
of individual events transmitting the infection during network-scale spreading dynamics.
This added level of realism further shortened the relay times. Despite of
the overestimation of the relay times when using the two first approximations, the overall picture arising from all the three
definitions remains the same; the dominant factor is burstiness.

As the timings of event sequences of individual links have been seen to be the most important factor for spreading dynamics, one might consider
mapping the dynamic problem to a static one by defining proper dynamic weights for the links of the aggregated network.
This route was recently taken by Miritello \emph{et al.}~\cite{Miritello10}, who defined the dynamic strength of a tie as the probability of it being able to transmit the infection during an SIR spreading process. For the special case of SI dynamics, this probability would always equal one -- hence, instead, for the SI process, the interesting quantity is the speed of propagation. One may thus define a dynamical weight that is inversely proportional to the speed \emph{e.g.}~as $n_t=T/\tauzero -1$ (or, including triggering effects, $n_t=T/\tauone -1$) where $n_t$ is the dynamic weight; for events with uniformly random timings (Eq.~\ref{eq:uniftimes_tauzero}), this weight would be equal to 
the number of events
on the link with some variation due to fluctuations. However, as we have seen with the actualized transfer times, the real fastest time-ordered paths taken by the spreading process may be considerably different \emph{e.g.}~as temporal shortcuts are employed, limiting the accuracy of this approximation.


\section{Acknowledgments}
  Financial support from EU's 7th Framework Program's FET-Open to
  ICTeCollective project no. 238597 and by the Academy of Finland, the
  Finnish Center of Excellence program 2006-2011, project no. 129670, and
  TEKES (FiDiPro) are gratefully acknowledged. We thank A.-L. Barab\'asi for the
  data used in this research.

\section{References}
\bibliography{long_spreading_main_intro}

\begin{thebibliography}{10}

\bibitem{TempReview}
P.~Holme and J.~Saram\"aki.
\newblock Temporal networks.
\newblock {\em arXiv: 1108.1780}, 2011.

\bibitem{Barrat2004}
A.~Barrat, M.~Barth\'{e}lemy, R.~Pastor-Satorras, and A.~Vespignani.
\newblock {The architecture of complex weighted networks}.
\newblock {\em Proceedings of the National Academy of Sciences of the United
  States of America}, 101(11):3747--3752, 2004.

\bibitem{Boccaletti06}
S.~Boccaletti, V.~Latora, Y.~Moreno, Chavez, and D.-U. Hwang.
\newblock Complex networks: Structure and dynamics.
\newblock {\em Phys. Rep.}, 424:175 –-- 308, 2006.

\bibitem{Vazquez07}
Alexei Vazquez, Bal\'azs R\'acz, Andr\'as Luk\'acs, and Albert-L\'aszl\'o
  Barab\'asi.
\newblock Impact of non-poissonian activity patterns on spreading processes.
\newblock {\em Phys. Rev. Lett.}, 98(15):158702, 2007.

\bibitem{Karsai11}
M.~Karsai, M.~Kivel\"a, R.~K. Pan, K.~Kaski, J.~Kert\'esz, A.-L. Barab\'asi,
  and J.~Saram\"aki.
\newblock Small but slow world: How network topology and burstiness slow down
  spreading.
\newblock {\em Phys. Rev. E}, 83(2):025102, 2011.

\bibitem{Miritello10}
Giovanna Miritello, Esteban Moro, and Rub\'{e}n Lara.
\newblock The dynamical strength of social ties in information spreading.
\newblock {\em Phys. Rev. E}, 83:045102, 2011.

\bibitem{Pan11}
R.~K. Pan and J.~Saram\"aki.
\newblock Path lengths, correlations, and centrality in temporal networks.
\newblock {\em Phys. Rev. E}, 84:016105, 2011.

\bibitem{Simonsen04}
Ingve Simonsen, Kasper~Astrup Eriksen, Sergei Maslov, and Kim Sneppen.
\newblock Diffusion on complex networks: a way to probe their large-scale
  topological structures.
\newblock {\em Physica A}, 336:163 -- 173, 2004.

\bibitem{Holme05}
Petter Holme.
\newblock Network reachability of real-world contact sequences.
\newblock {\em Phys. Rev. E}, 71(4):046119, 2005.

\bibitem{Barthelemy04}
Marc Barth\'elemy, Alain Barrat, Romualdo Pastor-Satorras, and Alessandro
  Vespignani.
\newblock Velocity and hierarchical spread of epidemic outbreaks in scale-free
  networks.
\newblock {\em Phys. Rev. Lett.}, 92(17):178701, 2004.

\bibitem{Onnela07}
J.~P. Onnela, J.~Saram\"aki, J.~Hyv\"onen, G.~Szab\'o, D.~Lazer, K.~Kaski,
  J.~Kert\'esz, and A.~L. Barab\'asi.
\newblock Structure and tie strengths in mobile communication networks.
\newblock {\em Proc. Natl. Acad. Sci. U.S.A.}, 104(18):7332--7336, 2007.

\bibitem{AndersonMay}
R.M. Anderson and May R.M.
\newblock {\em Infectious Diseases of Humans: Dynamics and Control}.
\newblock Oxford Science Publications, 1992.

\bibitem{Hethcote}
H.W. Hethcote.
\newblock The mathematics of infections diseases.
\newblock {\em SIAM Review}, 42:599, 2000.

\bibitem{Newman02}
M.E.J. Newman.
\newblock Spread of epidemic disease on networks.
\newblock {\em Phys. Rev. E}, 66:016128, 2002.

\bibitem{Kenah07}
Eben Kenah and James~M. Robins.
\newblock Second look at the spread of epidemics on networks.
\newblock {\em Phys. Rev. E}, 76:036113, 2007.

\bibitem{Morris97}
Martina Morris and Mirjam Kretzschmar.
\newblock Concurrent partnerships and the spread of hiv.
\newblock {\em AIDS}, 11(5):641—648, 1997.

\bibitem{RochaPNAS2010}
Luis E.~C. Rocha, Fredrik Liljeros, and Petter Holme.
\newblock {Information dynamics shape the sexual networks of Internet-mediated
  prostitution}.
\newblock {\em Proc. Natl. Acad. Sci. U.S.A.}, 107(13):5706--5711, 2010.

\bibitem{Rocha2011}
Luis E.~C. Rocha, Fredrik Liljeros, and Petter Holme.
\newblock Simulated epidemics in an empirical spatiotemporal network of 50,185
  sexual contacts.
\newblock {\em PLoS Comput Biol}, 7(3):e1001109, 2011.

\bibitem{Barabasi05}
A.~L. Barab\'asi.
\newblock The origin of bursts and heavy tails in human dynamics.
\newblock {\em Nature}, 435:207--211, 2005.

\bibitem{Iribarren10}
Jos\'{e}~Luis Iribarren and Esteban Moro.
\newblock Affinity paths and information diffusion in social networks.
\newblock {\em Social Networks}, 33:134--142, 2011.

\bibitem{Molloy95}
Michael Molloy and Bruce Reed.
\newblock A critical point for random graphs with a given degree sequence.
\newblock {\em Random Structures \& Algorithms}, 6:161–180, 1995.

\bibitem{Jo11}
H.-H. Jo, M.~Karsai, J.~Kert\'esz, and K.~Kaski.
\newblock Circadian pattern and burstiness in mobile phone communication.
\newblock {\em arXiv:1101.0377v2}, 2011.

\bibitem{Goh08}
K.-I. Goh and A.-L. Barab\'asi.
\newblock Burstiness and memory in complex systems.
\newblock {\em Europhys. Lett.}, 81:48002, 2008.

\bibitem{Wu10}
Ye~Wu, Changsong Zhou, Jinghua Xiao, J\"{u}rgen Kurths, and Hans~J.
  Schellnhuber.
\newblock Evidence for a bimodal distribution in human communication.
\newblock {\em Proc. Natl. Acad. Sci. U.S.A.}, 107(44):18803--18808, 2010.

\bibitem{Min11}
Byungjoon Min, K.-I. Goh, and Alexei Vazquez.
\newblock Spreading dynamics following bursty human activity patterns.
\newblock {\em Phys. Rev. E}, 83(3):036102, 2011.

\bibitem{Candia08}
Juli\'{a}n Candia, Marta~C Gonz\'{a}lez, Pu~Wang, Timothy Schoenharl, Greg
  Madey, and Albert-L\'{a}szl\'{o} Barab\'{a}si.
\newblock Uncovering individual and collective human dynamics from mobile phone
  records.
\newblock {\em J. Phys. A: Math. Theor.}, 41(22):224015, 2008.

\bibitem{Steutel67}
F.W. Steutel.
\newblock Random division of an interval.
\newblock {\em Statistical Neerlandica 21}, 21:231--244, 1967.

\bibitem{David03}
Herbert~A. David and H.N. Nagaraja.
\newblock {\em Order Statistics}.
\newblock Wiley-Blackwell, third edition, 2003.

\bibitem{Eisler08}
Z.~Eisler, I.~Bartos, and J.~Kert\'{e}sz.
\newblock Fluctuation scaling in complex systems: Taylor's law and beyond.
\newblock {\em Advances in Physics}, 57(1):89--142, 2008.

\bibitem{Krings09}
Gautier Krings, Francesco Calabrese, Carlo Ratti, and Vincent~D Blondel.
\newblock Urban gravity: a model for inter-city telecommunication flows.
\newblock {\em J. Stat. Mech.}, 2009(07):L07003, 2009.

\bibitem{Pan10}
R.~K. Pan, M.~Kivel\"a, J.~Saram\"aki, K.~Kaski, and J.~Kert\'esz.
\newblock Using explosive percolation in analysis of real-world networks.
\newblock {\em Phys. Rev. E}, 83:046112, 2011.

\bibitem{Moran47}
P.A.P. Moran.
\newblock The random division of an interval.
\newblock {\em Supplement to the Journal of the Royal Statistical Society},
  9:92--98, 1947.

\end{thebibliography}

\appendix

\section{Reference models and static and dynamic properties}
\label{sec:reference_models}

Empirical systems are usually characterized by properties which are extremely unlikely to be found in random systems. One way of studying static empirical networks is to compare them to an ensemble of graphs which possess some subset of the properties of the empirical system but which are otherwise maximally random. These ensembles are often called as \emph{reference} or \emph{null models}, and they can be used to perform ``controlled experiments'' to see which properties of the data are relevant for some observed phenomenon. The most convenient way of generating samples from random ensembles is to use shuffling methods which retain the selected set of features of the original system but randomize and destroy all other unrelated properties. Then, one can observe how the quantities of interest differ in the reference ensemble as compared to the empirical system. In this section, we define a reference model framework for temporal networks by formally defining the properties that the reference models modify. 

A temporal network can be described by a set of nodes $v \in V$ which are connected by a sequence of events, $\mathcal E = \{e_1,...,e_N\}$, occurring between the nodes during the observation period $[0,T]$. Each event can be defined as a quadruplet of the form $e \equiv (u,v,t,d)$, where $u$ and $v$ are the initiator and receiver of the event, $t \in [0,T]$ denotes the execution time and $d$ is the duration. In this study, we analyze mobile phone calls which have durations, but for simplicity we consider them instantaneous as their duration is irrelevant for the investigated processes. Consequently, the definition of an event is simplified to a triplet form $e \equiv (u,v,t)$. A sequence of such events can be used to construct an aggregated network $G(\mathcal E )$ where a link appears between nodes who participate in a common event and link weights can be defined as the total number of events $n$ taking place between connected nodes during the examined time period. 

Formally, a property of an event sequence can be any function $c$ that takes an event sequence as an argument and returns the property. Using this function the set of all possible event sequences can be divided into two subsets, one with sequences having the property, $C(\mathcal E)=\{\mathcal E'|c(\mathcal E) = c(\mathcal E') \}$ and the rest of the sequences that do not have the property. The reference model related to a property is an ensemble of event sequences containing all the sequences in $C(\mathcal E)$ where each sequence is equally probable. At the crudest level, the properties of an event sequence can be divided into static properties which can be detected from the aggregated network (\emph{i.e.} we can write $c(\mathcal E)=c'(G(\mathcal E))$) and to temporal properties for which we also need to consider the time stamps of the events. 

Let us first concentrate on the static properties. Without any constraints all of our event sequences have the same set of nodes $V$ and the ensembles of possible graphs induced by these sequences are limited to graphs with the same node set. The number of events in each sequence is $N$ and thus the number of edges in the graphs vary from $1$ to $N$. The first constraint we want to consider is the \emph{number of links} $c_{L}(\mathcal E)=L(G(\mathcal E))$, where $L$ is a function returning the number of of links in a graph. Now the ensemble of graphs induced by the ensemble of event sequences correspond to an ensemble of Erd\H{o}s-R\'enyi (E-R) random graphs with given numbers of nodes and edges. E-R graphs have a binomial degree distribution, instead of the fat-tailed distributions commonly observed in data. Consequently, the next step is to limit the graphs to have exactly the same \emph{degree sequence} as the empirical graph. That is, we define a property $c_{DD}(\mathcal E)=k(G( \mathcal E))$, where $k$ is the function returning the degree sequence of a graph. This ensemble is called the configuration model. The full topology of the network is returned if in addition to the degree sequence we include the topological configurations and restore the original connection structure. That is, we keep the property of \emph{topological configurations} with the constraint $c_C(\mathcal E)=A(G(\mathcal E))$, where $A$ is the unweighted adjacency matrix of the graph.

\begin{figure}[thl!]
  \begin{center}
     \includegraphics[width=9cm,angle=0]{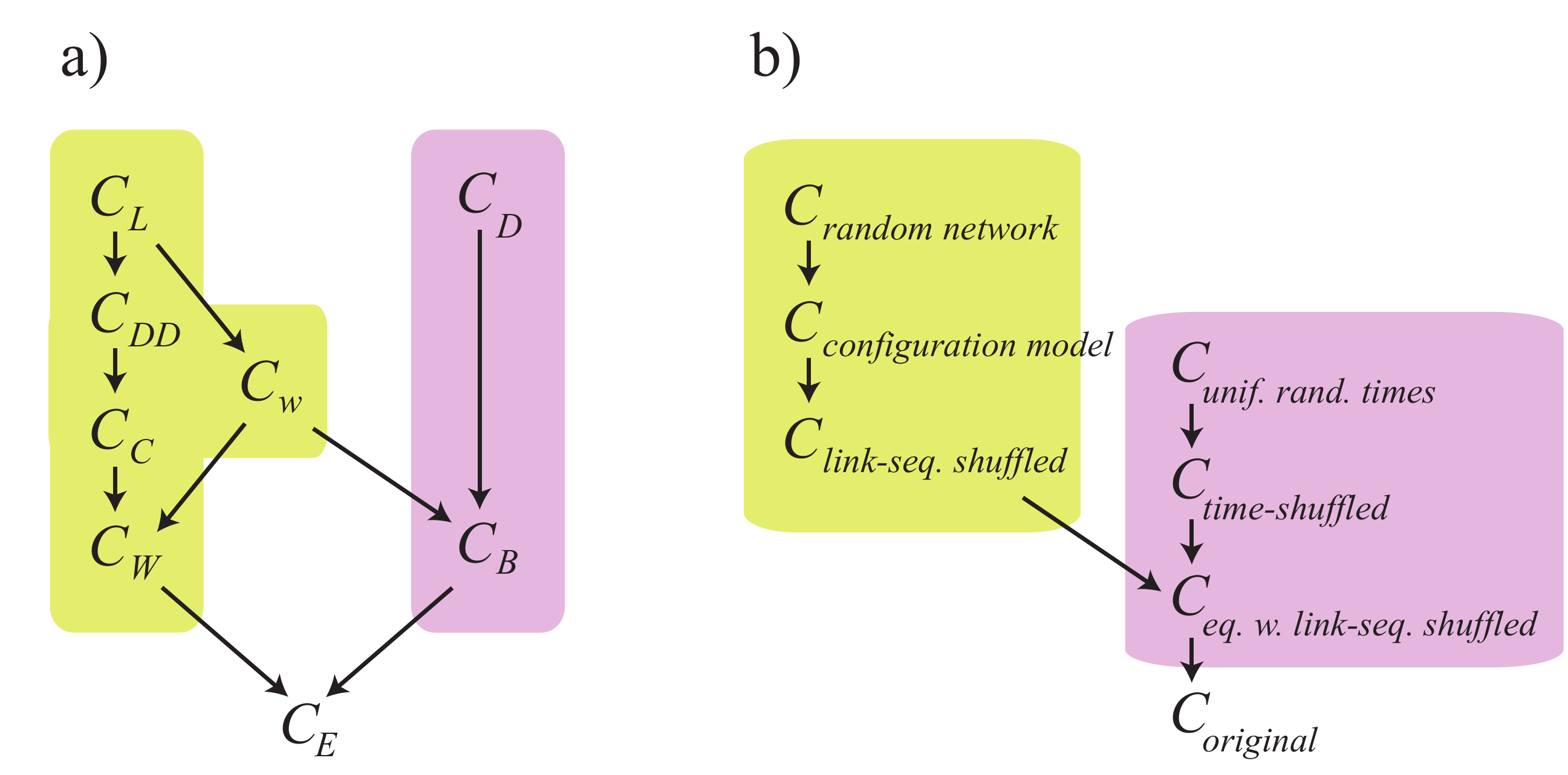}
  \end{center}
  \caption{Illustration of the hierarchical structure of (a) the properties discussed in the text, and (b) the randomized reference models. Arrows denote the inclusions of different ensembles in a way that there is an arrow from $A$ to $B$ if $A \supseteq B$. Static properties and reference models are located on the left-hand side, while temporal properties and reference models are on the right.}
   \label{fig:referenceModels}
\end{figure}

Until now, we have neglected the edge weights of the network, \emph{i.e.}~the number of events taking place between node pairs, and in our ensembles the weight distribution has been maximally random, or binomial. If we define link weights as the numbers of events on links, we can constrain the \emph{weight distribution} $c_w(\mathcal E)=N_w(G(\mathcal E))$, where
$N_w(G)(n)=\sum_{i,j} \delta(G_{i,j},n)$ is the count of weights in the network represented by the weight matrix $G$ and $n$ denotes the number of events on a link. Now the number of edges having a given weight is determined, but the weights are distributed randomly between edges. Finally, to restore all static properties, we can include information about which weight corresponds to which edge as we define the property corresponding to the \emph{weight-topology correlations} as the weight matrix of the aggregated graph, $c_W(\mathcal E)=G(\mathcal E)$. As we can conclude from the way we added increasingly strict constraints for the ensemble, the reference models present a hierarchical structure and the hierarchy relations can be expressed with the sets of event sequences having non-zero probability in the ensembles: $C_{L} \supseteq C_{DD} \supseteq C_{C} \supseteq C_W$ and $C_{L} \supseteq C_{w} \supseteq C_W$. This structure is also illustrated on Fig~\ref{fig:referenceModels} a).

\begin{figure}[thl!]
  \begin{center}
     \includegraphics[width=9cm,angle=0]{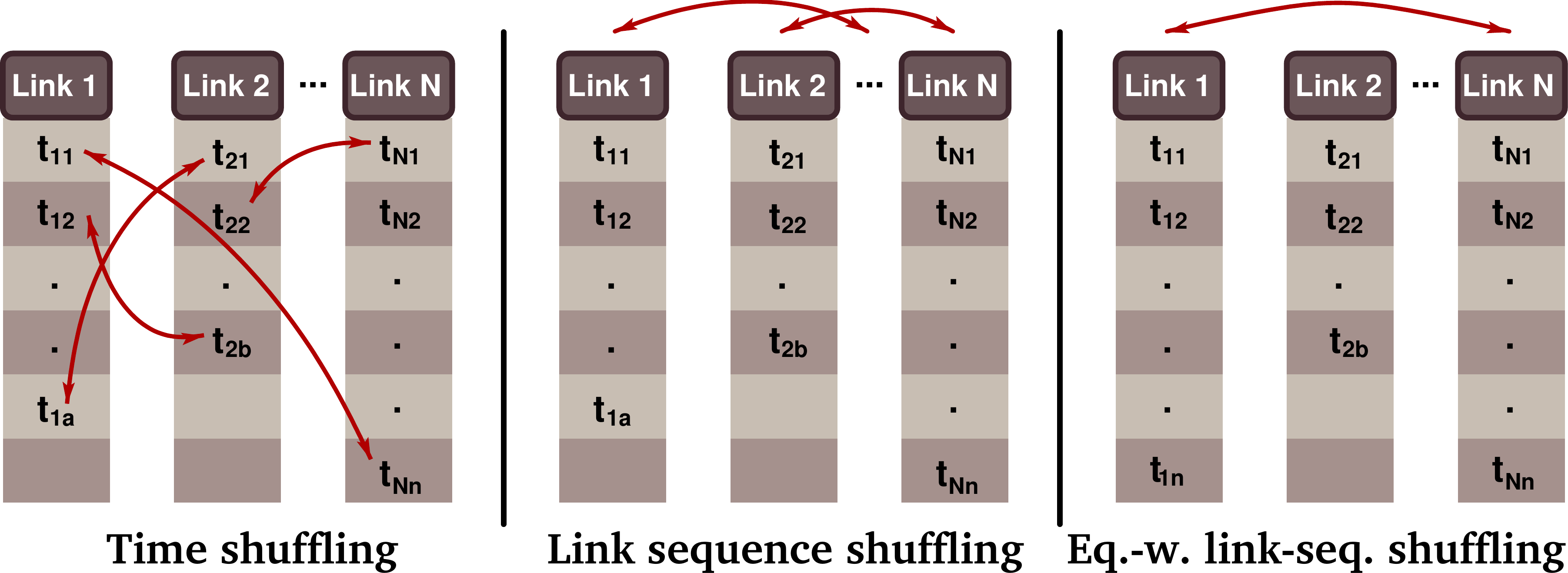}
  \end{center}
  \caption{Call sequence shuffling procedures corresponding to null models which only remove temporal inhomogeneities.}
   \label{fig:sequenceShuffling}
\end{figure}

The study of temporal inhomogeneities in networks has gained popularity only recently, and the related reference models are not well established. Next, we define three temporal properties which are used in this paper.  Without any temporal constraints, the times of the events in the ensembles are scattered uniformly at random on the interval $[0,T]$. First, we want to limit the ensemble in such a way that at least the network-wide frequency patterns in the event times are preserved.  To do so, we retain the \emph{global time stamp sequence} $c_D(\mathcal E)=\mathcal T$, where $\mathcal T=\{t_1,...,t_N\}$ is the sequence of the time stamps of the original event sequence $\mathcal E$. In a mobile phone network, the frequency patterns of the time stamps $\mathcal T$ reflect the daily and weekly fluctuations of the calling frequency, as on average people tend to place more calls at day time and less calls at night with a slight variation due to different weekdays. Going one step further, we would like to keep the distribution of \emph{the times of event sequences on links} $c_B(\mathcal E)= N_S(\mathcal E)$ with $N_S(\mathcal E)(s)=\sum_{i,j} \delta(S_{i,j},s)$, where $S_{i,j}=\{ t_{k_1},...,t_{k_n}|u_{k_1}=...=u_{k_n}=i,v_{k_1}=...=v_{k_n}=j\}$ is the sequence of times of events taking place between the nodes $i$ and $j$. The link sequences in the MPC network are known to be bursty. Restoring the full link sequences retains all the characteristics of the times of event sequences on links, including inter-event times and burstiness. The event sequences on links can be perceived as ``temporal link weights'', \emph{i.e.} sets of time stamps associated with each link instead of the simple scalar value of static weights. Similarly to static link weights,
 if in addition to retaining their overall distribution we want to keep the exact locations  of each link sequence in the network, then the complete original event sequence $\mathcal E$ is returned. Thus, for completeness we define a property of retaining the \emph{link sequence-topology correlations}  by $c_E(\mathcal E)=\mathcal E$. The inhomogeneities that are captured by $c_E$ (and not by any other property defined here) include burst-topology correlations and between-link correlations such as events causing other events in neighboring links, or for example some interlinked sets of nodes being more active than others at night.  For the temporal reference models, the hierarchy relations are $C_D \supseteq C_B \supseteq C_E$, and between temporal and static references they are $C_w \supseteq C_B, C_W \supseteq C_E$.

Any two properties $c_A$ and $c_B$ can be combined by defining that $c_{A+B}(\mathcal E)=\{c_{A}(\mathcal E),c_{B}(\mathcal E)\}$ or equivalently by having $C_{A+B}=C_A \cap C_B$. The different properties can be combined in any way to create reference models, but some combinations are redundant because the reference models can contain each other. All the reference models listed in the main text can now be defined as such combinations (see Sec.~\ref{sec:refmodels} and Table~\ref{table1}.)

\section{Role of boundary and initial conditions}
\label{sec:boundaryconditions}

\begin{figure}[th!] \centering
  \includegraphics[width=100mm]{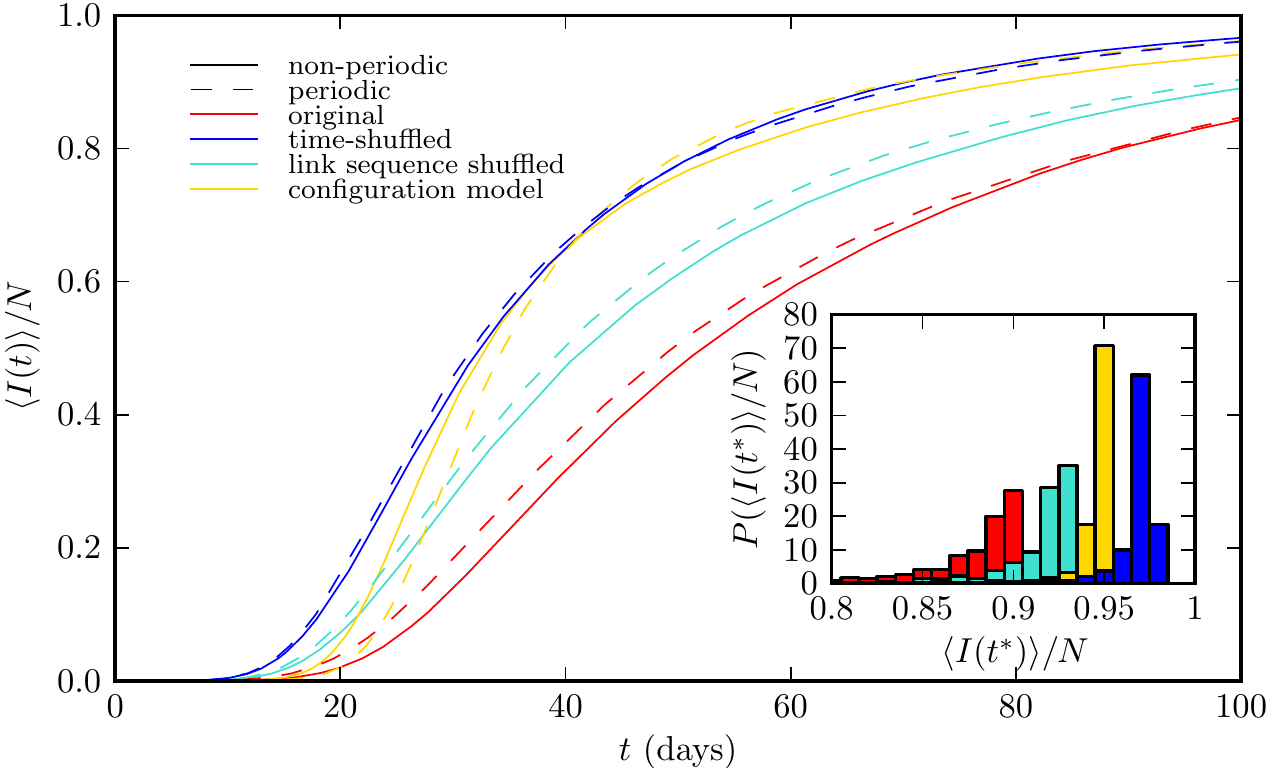}
\caption{
Comparison between spreading dynamics with (dashed line) and without (solid line) periodic temporal boundary conditions. Distribution of $\langle I(t^*) \rangle/N$ at time $t^*=100$ days of different null models without periodic boundary conditions is shown in the inset. 
}
\label{fig:boundary_conditions}
\end{figure}

To ensure that our results are not heavily affected by the periodic temporal boundary conditions that we applied, we have repeated some of our reference model calculations without periodic boundaries, using only the available finite time window of $120$ days. Here, the infection was initiated from random node at a random point in time within the first $20$ days, and the process was run for $100$ days. Repeating the measurements for $10^3$ times with different initial conditions, the average infection curves (solid lines in Fig.\ref{fig:boundary_conditions}) qualitatively show the same behavior as for the periodic boundary conditions (dashed lines in Fig.\ref{fig:boundary_conditions}). The overall effects are minimal, but the periodic boundary conditions slightly fasten the spreading process for the reference models having the original link-sequence distribution. For the time-shuffled event sequence where bursty temporal correlations do not play a role, the two curves (with and without periodic boundaries) overlap almost perfectly. Since without periodic boundary conditions the system does not reach full prevalence during the examined time period, we recorded the distribution of the infection curve values $\langle I(t=100)\rangle /N$ at the end of the time window (see Fig.\ref{fig:boundary_conditions} inset) in order to see the effects of random initial conditions. Although there is variance and the tails of the distributions overlap, their order is the clearly same as for the means of the curves.

\begin{figure}[ht!] \centering
  \includegraphics[width=100mm]{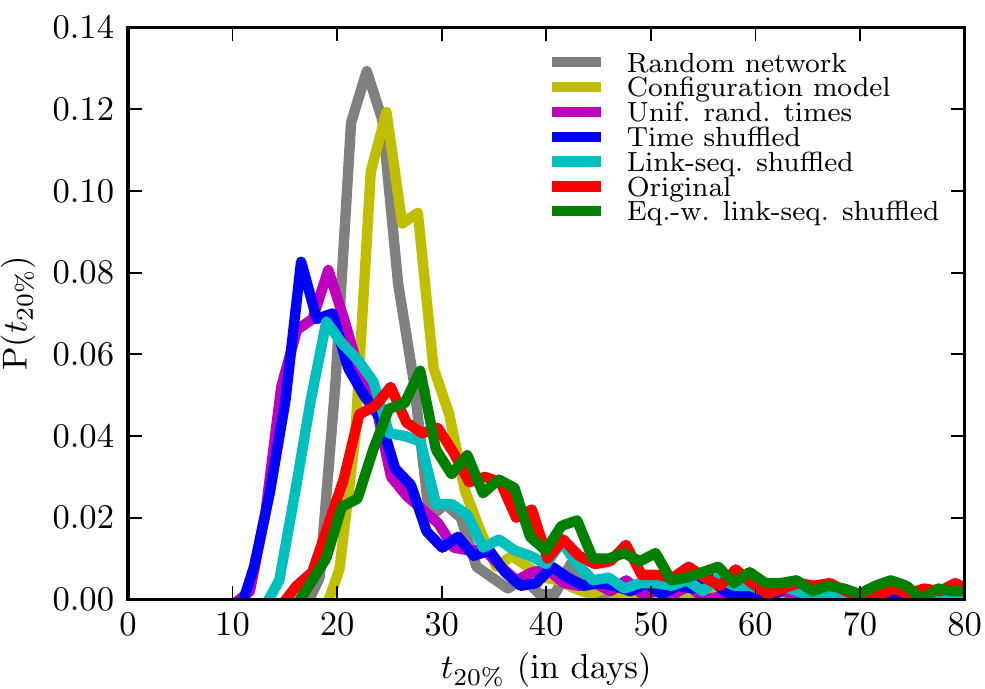}
\caption{Distributions for times to reach $20\%$ prevalence.}
\label{fig3}
\end{figure}

The initial conditions, \emph{i.e.}~the first infected node and the starting time, can have a large impact on the spreading process. The topological position and the activity of the initially infected node strongly influences the speed of a realization of the process, as the infection lifts off faster when it is initiated from a well-connected hub with a high level of activity. Such differences are reflected in the variation of the distributions of the times to reach prevalence levels of  $20\%$ and $100\%$, as shown in Fig.\ref{fig3} and in the inset of Fig.\ref{fig:SI_plots}, respectively. The effects of the initial conditions can be quantified by the width of the distributions, as summarized in Table \ref{table3}, where the average and the error of the mean values of the distributions of times to reach prevalences of $20\%$ and $100\%$ are shown for all models. Especially for early-stage distributions in Fig.\ref{fig3}, the reference models that retain more inhomogeneities have a larger spread.

\begin{table}[h!]
\begin{center}
\begin{tabular}{|l||c|c|}
\hline
EVENT SEQUENCE & $\langle t_{20\%} \rangle$ & $\langle t_{100\%} \rangle$\\ \hline\hline
Original & $38.242 \pm 0.741$ & $677.670 \pm 1.008$\\
\hline
Eq.-w. link-seq. shuffled & $41.206 \pm 0.831$ & $597.267 \pm 1.592$\\
\hline
Time shuffled & $26.486 \pm 0.527$ & $402.400 \pm 1.141$\\
\hline
Unif. rand. times & $24.962 \pm 0.390$ & $401.649 \pm 1.113$\\
\hline
Link-seq. shuffled & $30.676 \pm 0.552$ & $495.981 \pm 1.002$\\
\hline
Configuration model & $29.359 \pm 0.327$ & $446.621 \pm 0.949$\\
\hline
Random network & $26.643 \pm 0.291$ & $411.803 \pm 1.139$\\
\hline
\end{tabular}
\end{center}
\caption{Average time for reaching prevalence levels of  $20\%$ and $100\%$ infection, and error of the mean values for each model.}
\label{table3}
\end{table}

\begin{figure*}[th!] \centering
\subfigure[Original event sequence]{
  \includegraphics[width=56mm]{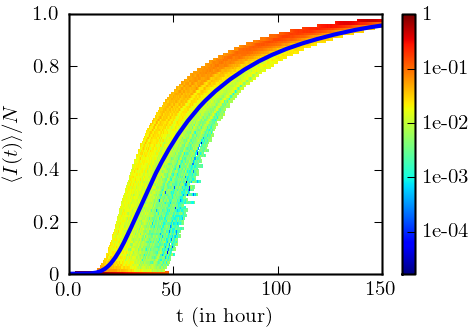}
}
\subfigure[Time shuffled event sequence]{
  \includegraphics[width=56mm]{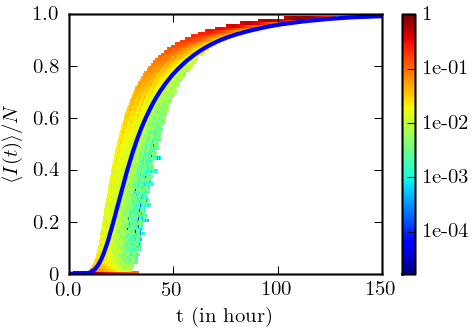}
}
\subfigure[Configuration model event sequence]{
  \includegraphics[width=56mm]{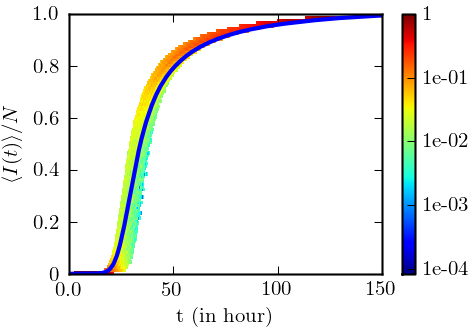}
}\\
\caption{Density and average of infection curves in the time-infection space for different null models. (a) Original event sequence, (b) time shuffled event sequence and (c) configuration model. On each figure the $75\%$ quantiled distributions are shown together with the average infection curves (solid line).}
\label{fig:5}
\end{figure*}

A more sophisticated picture can be drawn by recording the density of the infection curves in a time-prevalence space. For computing the density, this space
can be divided into  bins of equal size; then, the number of times when an infection curve passed through it can be counted for each bin cell. In Fig.\ref{fig:5} we show the normalized density figures of three selected reference models together with the average curves. In order to display the regime where the vast majority of the infection curves are running, we have used quantiles for the distributions for each row and show only the $75\%$ of the curves symmetrical around the median. This demonstrates that even though initial conditions cause an impact on the spreading evolution, most of the curves are not too far from the ensemble average which gives then a fairly good estimate for the actual behaviour. The effect of heterogeneities increasing the spread  is also visible here, as the dispersion of the individual curves is decreasing around the average and the $75\%$ quantile area becomes more narrow for reference models with less constraints.

\section{Network size dependence}
\label{sec:citynetworks}

In order to study the dependence of the speed of SI spreading process on
the system size, we need to extract subnetworks of the MPC network. Here,
we take use of the fact that MPC
networks display geographical embedding~\cite{Krings09}, and 
divide the users into subsets by considering in which city they live
according to the postal code of their subscriptions~\cite{Pan10}. We then
determine the largest connected component (LCC) of each city network 
and construct
the event sequence data of that city by selecting all the calls between the
users in the LCC. The above procedure yields 7569
subnetworks, whose sizes vary over five orders of magnitude. Note that the
information about the postal code of subscription is available for only
about half of the users, and hence the amount of users in all
subnetworks is less than in the entire MPC network.

MPC networks inside cities form natural subsystems of the 
whole MPC network \cite{Pan10,Krings09}. There is
no reason to believe that the social systems of different sizes are similar; 
instead, it might be possible that the
interaction patterns of people in the entire country might have very
different characteristics as compared to the interaction patterns inside a
small city. Further, the size of the city might also have some effect on the
patterns and correlations of interaction between users.  We start by
inspecting the properties of the corresponding 
static networks.

\begin{figure}
  \begin{center}
    \includegraphics[width=0.7\linewidth]{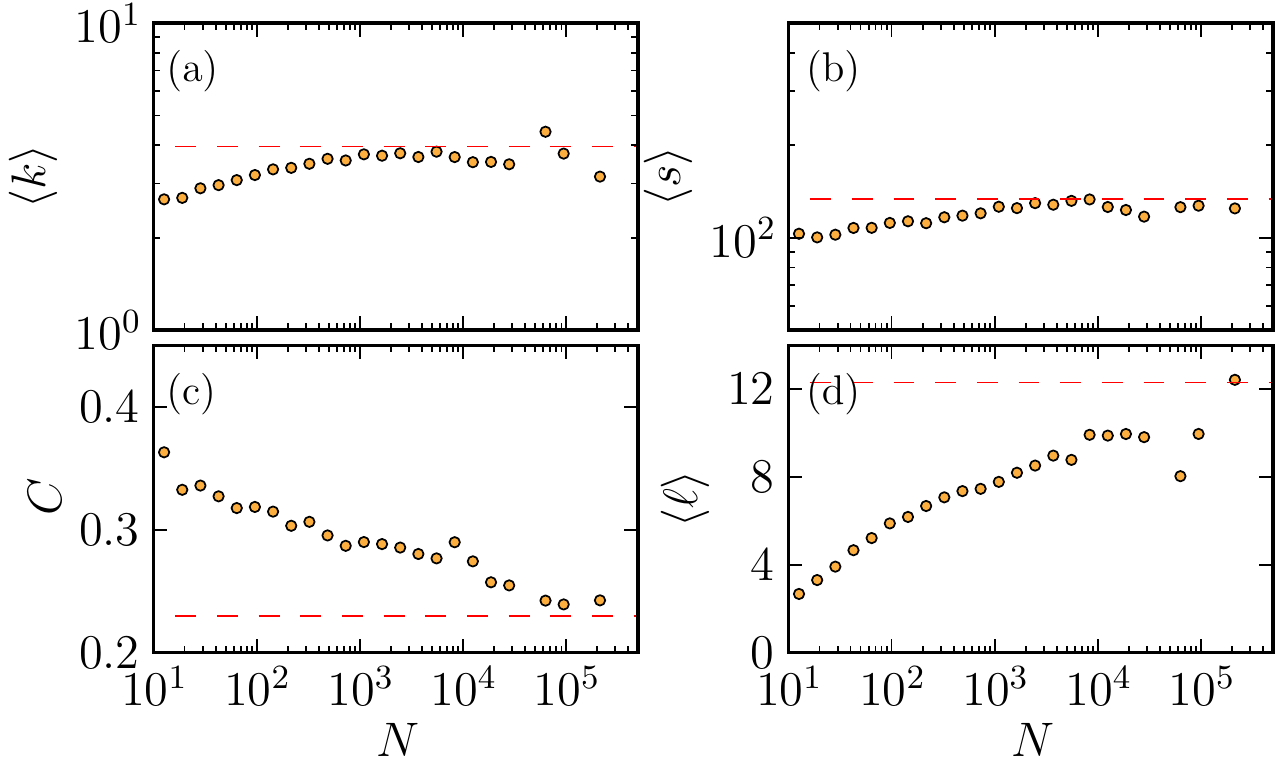}
  \end{center}
  \caption{Variation of the static properties of the subnetworks with the
  system size, $N$. Plot of the (a) average degree $\langle k \rangle$, (b)
  average distance $\langle \ell \rangle$, (c) average strength $\langle s
  \rangle$, and the (d) clustering coefficient $C$, of the subnetworks, as
  a function of their size. The dashed line represents the corresponding
  value for the entire MPC network.}
  \label{fig:cityStatic}
\end{figure}

The average degree $\langle k \rangle$ of the subnetworks initially
increases with  $N$, but for larger systems it saturates when the value for the entire MPC network is reached
[Fig.~\ref{fig:cityStatic}~(a)].  The average strength $\langle s \rangle$,
i.e. the average number of calls in which nodes participate,  
also behaves similarly~[Fig.~\ref{fig:cityStatic}~(b)].  These
properties suggest that although there is a large variation in the sizes
of the subnetworks, their fundamental properties are similar. Typical to
social networks, the clustering coefficients $C$ of the entire MPC network
and the subnetworks are much higher than would be expected from Erd\H{o}s-R\'enyi random
networks with corresponding size and average
degree~[Fig.~\ref{fig:cityStatic}~(c)]. The value of $C$ decreases with increasing
network size, reaching the value of the entire MPC network
for large $N$. Fig.~\ref{fig:cityStatic}~(d), shows the variation of the
average shortest path length in the subnetworks
$\ell$ with $N$. As
expected, we find that the \emph{static} MPC network is a small-world
system where (roughly) $\langle \ell \rangle \propto \ln N$. 
Note that when  $N>2\times10^3$, we
calculate $\langle \ell \rangle$ by sampling the distance from $10^3$
randomly chosen nodes to all other nodes in the network.

\begin{figure}
  \begin{center}
    \includegraphics[width=7.5 cm]{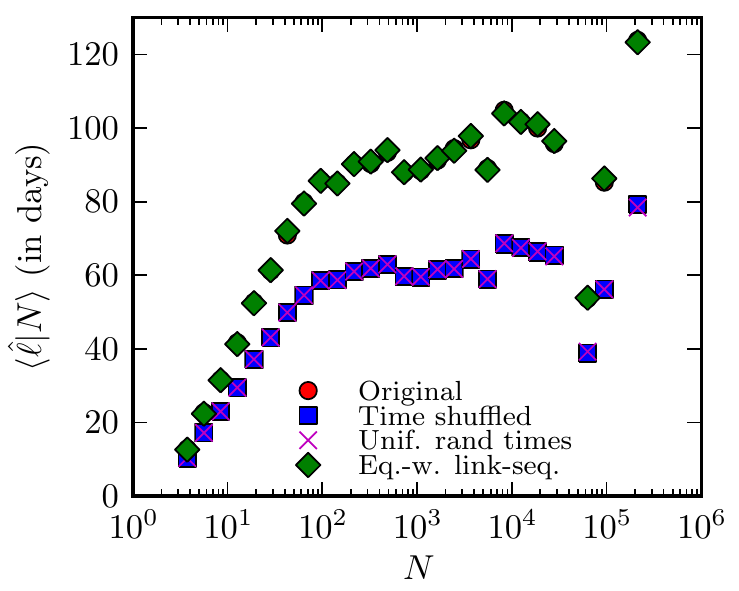}
  \end{center} 
  \caption{The
  network-wide average temporal distance between two nodes (time to reach $i$ from $j$) as a function of
  the number of nodes $N$ in the network for the original data and several null models. 
  The original curve overlaps with the equal-weight link-sequence shuffled model;
  the time-shuffled and uniform random times curves also practically overlap.
  All curves are
  averaged over $10^3$ initial conditions.}
  \label{fig:globalSpreadingAux}
\end{figure}

Next, we perform the SI spreading process for each of the subsystems, using
both the original event sequence and the temporal reference
models. We simulate the process until all the nodes in the system are
infected and observe the infection time of each node. We then
measure the average time at which nodes get infected in a city as a
function of the city size $N$ [Fig~\ref{fig:globalSpreadingAux}]. This
average infection time is same as the average temporal distance (duration of 
the fastest paths) of the system. The temporal distances first show
a fast increase, followed by slower increase. Importantly, 
for all system sizes, the temporal distances
for the original data and the null models show a similar relationship. This indicates
that the observed results are unlikely to be an artifact associated with the
specific system, that is, the full MPC network.

\section{$\tauzero$ for the uniformly random times reference model with periodic boundary conditions}
\label{sec:order_statistics}
In this Appendix we are going to calculate the inter-event time and random
relay time distributions for the uniformly random times reference model. Further,
we are going to derive formulas for the expected average relay time and
variance for the average relay time.

In the uniformly random times reference model the sequences of events of  
edges do not depend on each other and we can calculate all statistics
just by considering a single edge at a time.
That is, we have an edge with $n$ events, each taking place 
uniformly at random in the interval $[0,T]$, and denote the time of the $i$th event (ordered by time)
by $t_i$.

The time of $i$-th event is distributed like the $i$-th order statistics of the uniform probability
distribution $t_i/T \sim U_{(i)}$. That is, $t_i/T$ follows the beta distribution \cite{David03}
\begin{equation}
p(t_i)=\frac{(t_i/T)^{i-2}(1-t_i/T)^{n-i+1}}{B(i-1,n-i)},
\label{eq:unif_t_dist}
\end{equation}
where $B(i-1,n-i)=\int_0^1 x^{i-2}(1-x)^{n-i+1} dx$ is the beta function.

In order to be able to calculate $\tauzero$, we need the distribution for the
inter-event times $\tau_i$, rather than the times of the events. With periodic temporal boundary
conditions in place, we have
\begin{equation}
\tau_i=
\begin{cases}
T-t_{n}+t_1 & \text{if } i=1 \\
t_{i}-t_{i-1} & \text{if } i\neq 1
\end{cases}.
\label{eq:iet_with_os}
\end{equation}
Note that the values of $t_i$ are not independent of each other. 
However, from Ref.~\cite{David03}, we know that if we
define a new random variable $W_{rs}=U_{(s)}-U_{(r)}$ we get
\begin{equation}
p_{W_{rs}}(w)=\frac{w^{s-r-1}(1-w)^{n-s+r}}{B(s-r,n-s+r+1)}
\label{eq:w_dist}
\end{equation}
Applying Eq.~\ref{eq:w_dist} to Eq.~\ref{eq:iet_with_os}, the inter-event time $\tau_i$ distribution takes the form
\begin{equation}
p(\tau_i)=
\begin{cases}
\frac{(\tau_i/T)^{n-2}(1-\tau_i/T)}{B(n-1,2)} & \text{if } i=1\\
\frac{(1-\tau_i/T)^{n-1}}{B(1,n)}& \text{if } i\neq 1
\end{cases}.
\label{eq:unif_tau_dist}
\end{equation}
The expected value of $\tau_i$ thus becomes
\begin{equation}
\ensavg{\tau_i}=
\begin{cases}
\frac{2T}{n+1}& \text{if } i=1\\
\frac{T}{n+1}& \text{if } i\neq 1
\end{cases},
\label{eq:unif_tau_exp}
\end{equation}
and the variance is
\begin{equation}
\text{Var}(\tau_i)=
\begin{cases}
\frac{2(n-1)}{(n+2)(n+1)^2}T^2& \text{if } i=1\\
\frac{n}{(n+2)(n+1)^2}T^2& \text{if } i\neq 1
\end{cases}.
\label{eq:unif_tau_var}
\end{equation}
Note that $\tau_1$ is longer than all the other inter-event times. At first this might seem to be a bias
in the reference model. However, if one would change the labeling scheme of the event times such that
$t_1$ would be the first event after some point $x \in [0,1]$ and use the periodic boundary condition
in labeling all other events such that $t_{n-1}$ would be the last event before $x$, this would not change the distribution
of $\tau_i$. This might seem paradoxical, but it just reflects the fact that if we select any point $x$ in the time
interval, we are more likely to hit long inter-event times than short inter-event times. That is, selecting the
point $x=0$ as basis of our ``coordinate system'' causes the first inter-event time to be longer than the others. 

Equipped with the functional form of the probability density function of the distribution of $\tau_i$, we can calculate 
the expected value of $\tauzero$ by using the
Eq.~\ref{tau0_iee}:
\begin{equation}
\ensavg{\avg{\tauzero}} =\ensavg{\frac{\sum_{i=1}^{n}\iet_i^2}{2T}}=\frac{\sum_{i=1}^{n}\ensavg{\iet_i^2}}{2T}=\frac{\ensavg{\iet_1^2}+(n-1)\ensavg{\iet_2^2}}{2T}.
\label{eq:unif_tau0_iee}
\end{equation}
Now noting that $\ensavg{\iet_i^2}=\ensavg{\iet_i}^2+\text{Var}(\iet_i)$ and plugging in Eq~\ref{eq:unif_tau_exp} and Eq.~\ref{eq:unif_tau_var}
we get
\begin{equation}
\ensavg{\avg{\tauzero}}=\frac{T}{n+1}.
\label{eq:expectedtauzero}
\end{equation}

The calculations can be made easier if we only want to calculate the average relay times and do not care about 
the distribution of $\iet_i$. We can choose a random event $i$ and set our
 time reference in a way that $t_i=0$ and still apply the periodic boundary conditions. The 
set of inter-event times is retained in this transformation, and only the order of the times is lost. Further, the other $n-1$ events
(not including event $i$) are uniformly randomly distributed over the time interval $[0,T]$.
Thus, the problem is of finding the distribution of the average relay time can be solved by solving the problem 
of dividing the unit interval into $n$ intervals by $n-1$ uniformly random points. In Ref.~\cite{Moran47} the $l$'th moment
of the sum of the squares of the interval sizes was given as
\begin{equation}
\langle (\sum_{i=1}^n (\tau_i/T)^2)^l \rangle=\frac{(n-1)!l!}{(n+2l-1)!}\sum_{j_1+...+j_n=l}\frac{(2j_1)!...(2j_n)!}{j_1!...j_n!}.
\label{eq:unitintervalmoments}
\end{equation} 
Now the $l$'th moment of the average relay time can be calculated by using Eq.~\ref{eq:unitintervalmoments} with
\begin{equation}
\langle \overline{\tau_R}^l \rangle = (\frac{T}{2})^l \langle (\sum_{i=1}^n (\tau_i/T)^2)^l \rangle .
\label{eq:tauzeromoments}
\end{equation}

Using Eqs.~\ref{eq:unitintervalmoments} and \ref{eq:tauzeromoments} we obtain the correct average relay
time of  Eq.~\ref{eq:expectedtauzero}, and values for the second moment
\begin{equation}
\langle \overline{\tau_R}^2 \rangle = \frac{n+5}{(n+3)(n+2)(n+1)} T^2,
\end{equation}
and the variance
\begin{equation}
\text{Var} (\overline{\tau_R}) = \frac{n-1}{(n+1)^2(n+2)(n+3)}T^2.
\end{equation}

\end{document}